\documentclass[aps,prl,reprint,superscriptaddress]{revtex4-1}

\usepackage{amssymb,amsmath,amsthm,amsfonts,dsfont}
\usepackage{natbib}
\setcitestyle{numbers,comma,sort&compress}
\usepackage[colorlinks]{hyperref}
\usepackage[all]{hypcap}
\usepackage{url}
\usepackage[font=small]{caption}
\usepackage[labelformat=simple]{subcaption}
\usepackage{float}
\usepackage{algorithmic}
\usepackage{enumitem}
\usepackage[margin=1in]{geometry}

\usepackage{xcolor}
\usepackage{pgfplots}

\usepackage{mathtools}
\mathtoolsset{centercolon}

\DeclarePairedDelimiterX\braket[2]{\langle}{\rangle}{#1 \delimsize\vert #2}

\newcommand{\eq}[1]{(\ref{eq:#1})}
\newcommand{\thm}[1]{\hyperref[thm:#1]{Theorem~\ref*{thm:#1}}}
\newcommand{\defn}[1]{\hyperref[defn:#1]{Definition~\ref*{defn:#1}}}
\newcommand{\lem}[1]{\hyperref[lem:#1]{Lemma~\ref*{lem:#1}}}
\newcommand{\prop}[1]{\hyperref[prop:#1]{Proposition~\ref*{prop:#1}}}
\newcommand{\fig}[1]{\hyperref[fig:#1]{Figure~\ref*{fig:#1}}}
\newcommand{\tab}[1]{\hyperref[tab:#1]{Table~\ref*{tab:#1}}}
\renewcommand{\sec}[1]{\hyperref[sec:#1]{Section~\ref*{sec:#1}}}
\newcommand{\append}[1]{\hyperref[append:#1]{Appendix~\ref*{append:#1}}}
\newcommand{\cor}[1]{\hyperref[cor:#1]{Corollary~\ref*{cor:#1}}}

\newcommand{\norm}[1]{\left\lVert#1\right\rVert}

\pgfplotsset{
	log x ticks with fixed point/.style={
		xticklabel={
			\pgfkeys{/pgf/fpu=true}
			\pgfmathparse{exp(\tick)}%
			\pgfmathprintnumber[fixed relative, precision=3]{\pgfmathresult}
			\pgfkeys{/pgf/fpu=false}
		}
	},
	log y ticks with fixed point/.style={
		yticklabel={
			\pgfkeys{/pgf/fpu=true}
			\pgfmathparse{exp(\tick)}%
			\pgfmathprintnumber[fixed relative, precision=3]{\pgfmathresult}
			\pgfkeys{/pgf/fpu=false}
		}
	}
}

\usepackage{ mathrsfs }

\newcommand{\scrS}{\mathscr{S}}
\newcommand{\scrT}{\mathscr{T}}

\usepackage{pdfpages}
\usepackage{pgffor}
\makeatletter
\AtBeginDocument{\let\LS@rot\@undefined}
\makeatother

\begin{document}


\title{Nearly optimal lattice simulation by product formulas}

\author
{Andrew M.\ Childs and Yuan Su\\
	\textit{\normalsize{
	Department of Computer Science,
	Institute for Advanced Computer Studies, and
	Joint Center for Quantum Information and Computer Science,
	University of Maryland}}
}

\date{\today}

\begin{abstract}
We consider simulating an $n$-qubit Hamiltonian with nearest-neighbor interactions evolving for time $t$ on a quantum computer. We show that this simulation has gate complexity $(nt)^{1+o(1)}$ using product formulas, a straightforward approach that has been demonstrated by several experimental groups. While it is reasonable to expect this complexity---in particular, this was claimed without rigorous justification by Jordan, Lee, and Preskill---we are not aware of a straightforward proof. Our approach is based on an analysis of the local error structure of product formulas, as introduced by Descombes and Thalhammer and further simplified here. We prove error bounds for canonical product formulas, which include well-known constructions such as the Lie-Trotter-Suzuki formulas. We also develop a local error representation for time-dependent Hamiltonian simulation, and we discuss generalizations to periodic boundary conditions, constant-range interactions, and higher dimensions. Combined with a previous lower bound, our result implies that product formulas can simulate lattice Hamiltonians with nearly optimal gate complexity.
\end{abstract}


\maketitle

\noindent
Simulating the Hamiltonian dynamics of a quantum system is one of the most natural applications of a quantum computer. Indeed, the idea of quantum computation, as suggested by Feynman \cite{Fey82} and others, was primarily motivated by the problem of quantum simulation. Quantum computers can simulate a variety of physical systems, including quantum chemistry \cite{BMWAW15,WBCHT14,Pou15,Cao19}, quantum field theory \cite{JLP12,JLP14a}, and many-body \mbox{physics \cite{RWS12}}, and could ultimately lead to practical applications such as designing new pharmaceuticals, catalysts, and materials \cite{BWMMNC18,deWolf2017}.

A natural class of Hamiltonians that includes many physically reasonable systems is the class of lattice Hamiltonians \cite{JLP12,LLA15,HHKL18,Gao17}. Lattice Hamiltonians arise in many models of condensed matter physics, including systems of spins (e.g., Ising, XY, and Heisenberg models; Kitaev's toric code and honeycomb models; etc.), fermions (e.g., the Hubbard model and the $t$-$J$ model), and bosons (e.g., the Bose-Hubbard model). Note that fermion models can be simulated using local interactions among qubits by using a mapping to qubits that preserves locality \cite{VC05}. Digital simulations of quantum field theory also typically involve approximation by a lattice system \cite{JLP12}.

For simplicity, we mainly focus on nearest-neighbor lattice systems in one dimension (although we discuss generalizations to other lattice models as well). In this case, $n$ qubits are laid out on a one-dimensional lattice and the Hamiltonian only involves nearest-neighbor interactions. Specifically, a Hamiltonian $H$ is a lattice Hamiltonian if it acts on $n$ qubits and can be decomposed as $H=\sum_{j=1}^{n-1}H_{j,j+1}$, where each $H_{j,j+1}$ is a Hermitian operator that acts nontrivially only on qubits $j$ and $j+1$. We assume that $\max_j\norm{H_{j,j+1}}\leq1$, for otherwise we evolve under the normalized Hamiltonian $H/\max_j\norm{H_{j,j+1}}$ for time $\max_j\norm{H_{j,j+1}}t$.

Lloyd's original proposal for an explicit quantum simulation algorithm \cite{Llo96} uses the Lie-Trotter product formula. Subsequent work achieves better asymptotic complexity \cite{BACS05} using higher-order Suzuki formulas \cite{Suz91}. We refer to all such formulas as \emph{product formulas}. The product-formula algorithm is straightforward yet surprisingly efficient for quantum simulation. Indeed, it can conserve certain symmetries of the dynamics \cite{HS05} and appears to be advantageous for various practical systems \cite{CMNRS18,BMWAW15,RWSWT17}. Although recent simulation algorithms have better asymptotic complexities \cite{BCK15,BC12,BCCKS14,LC16,LC17,Low18,BCG14,Campbell18,Kieferova18,LW18}, the product-formula approach remains a natural choice for experimental simulations \cite{BCC06,Bar15,Lan11} due to its simplicity and the fact that it does not require any ancilla qubits. Its study has also illuminated areas beyond quantum computing \cite{bk:BC16}.

One of the main challenges in quantum simulation is to analyze the gate complexity of simulation algorithms. Explicit gate counts are especially desirable for near-term simulation because early quantum computers will only be able to reliably perform a limited number of gates. While existing analysis appears to be tight for recent simulation algorithms, the product-formula bound can be loose by several orders of magnitude \cite{CMNRS18,RWS12,RWSWT17,BMWAW15}. This dramatic gap makes it hard to identify the fastest simulation algorithm and to find optimized implementations for near-term applications \cite{CMNRS18}.

Product formulas can simulate a lattice system with fixed accuracy with gate complexity $O\big(n(nt)^2\big)$ in the first-order case and $O\big(n(nt)^{1+\frac{1}{2k}}\big)$ in the ($2k$)th-order case. However, it is natural to expect a more efficient simulation. Roughly speaking, a system simulates its own evolution for constant time using only constant circuit depth---and hence an extensive number of gates---so one might expect a true simulation complexity of $O(nt)$. Indeed, Jordan, Lee, and Preskill claimed that product formulas can simulate an $n$-qubit lattice system with $(nt)^{1+o(1)}$ gates \cite{JLP12}, but they did not provide rigorous justification and it is unclear how to formalize their argument. Subsequent work improves the analysis of the product-formula algorithm using information about commutation among terms in the Hamiltonian \cite{Somma16,CMNRS18,Pou15,WBCHT14}, the distribution of norms of terms \cite{HP18}, and by randomizing the ordering of terms \cite{COS18,Zha10}. However, none of these improvements can achieve the claimed gate complexity $(nt)^{1+o(1)}$ for lattice simulation.

\medskip
\noindent\textbf{Main result.}
Let $H=\sum_{j=1}^{n-1}H_{j,j+1}$ be an $n$-qubit lattice Hamiltonian. We order the terms in the \emph{even-odd} pattern $H_{1,2},H_{3,4},\ldots,H_{2,3},H_{4,5},\ldots$ obtaining the first-order product formula
\begin{equation}
\begin{aligned}
\label{eq:pf1}
	\scrS_1(t):&=\prod_{k=1}^{\frac{n}{2}-1}e^{-it H_{2k,2k+1}}\prod_{k=1}^{\frac{n}{2}}e^{-itH_{2k-1,2k}}\\
	&=e^{-itH_{\text{even}}}e^{-itH_{\text{odd}}}
\end{aligned}
\end{equation}
and the ($2k$)th-order product formulas
\begin{equation}
\begin{aligned}
\label{eq:pf2k}
\scrS_{2}(t)&:=e^{-i\frac{t}{2}H_{\text{odd}}}e^{-itH_{\text{even}}}e^{-i\frac{t}{2}H_{\text{odd}}}\\
\scrS_{2k}(t)&:=\scrS_{2k-2}(p_{k}t)^2 \, \scrS_{2k-2}((1-4p_{k})t) \, \scrS_{2k-2}(p_{k}t)^2
\end{aligned}
\end{equation}
with $p_{k}:=1/(4-4^{1/(2k-1)})$. Our main result is an asymptotic upper bound on the product-formula error
\begin{equation}
\begin{aligned}
\label{eq:main_result}
	\norm{\scrS_1(t)-e^{-itH}}&=O(nt^2)\\
	\norm{\scrS_{2k}(t)-e^{-itH}}&=O(nt^{2k+1}),
\end{aligned}
\end{equation}
where $\norm{\cdot}$ denotes the spectral norm.

The above error bound works well only for very small $t$. To simulate for a longer time, we divide the entire evolution into $r$ segments, and within each segment, we simulate using product formulas. To achieve accuracy $\epsilon$, it suffices to choose $r_1=O(nt^2/\epsilon)$ for the first-order formula and $r_{2k}=O\big(t(nt/\epsilon)^{\frac{1}{2k}}\big)$ for the ($2k$)th-order formula. Equivalently, we have gate complexity $O\big((nt)^2\big)$ and $O\big((nt)^{1+\frac{1}{2k}}\big)$ for the first- and ($2k$)th-order algorithm, assuming that we simulate with constant accuracy.

For any $\delta>0$, we choose an integer $k$ sufficiently large so that $\frac{1}{2k}\leq\delta$, upper-bounding the gate complexity as $O\big((nt)^{1+\frac{1}{2k}}\big)= O\big((nt)^{1+\delta}\big)$. This proves that the product-formula algorithm has asymptotic gate complexity $(nt)^{1+o(1)}$. Combining with the lower bound of $\widetilde{\Omega}(nt)$ established in \cite{HHKL18}, we have showed that product formulas can simulate a lattice Hamiltonian with nearly optimal gate complexity.

\medskip
\noindent\textbf{Applications.}
As an immediate application, our result gives a rigorous proof of the Jordan-Lee-Preskill claim about the complexity of simulating quantum field theory \cite{JLP12}. Recent works have analyzed the gate complexity of other quantum field theory simulations \cite{Preskill18}, including digital simulation of gauge theories \cite{LLY19}. The lattice Hamiltonians there have similar locality, so our analysis still applies. We expect our technique can be generalized to speed up the simulation of other systems, such as electronic structure Hamiltonians \cite{BWMMNC18}, power-law decaying interactions \cite{Tran18}, exponentially decaying interactions \cite{ME11}, and clustered Hamiltonians \cite{PHOW19}, but we leave a thorough study of such generalizations as a subject for future work \cite{CSTWWZ19}.

To simulate an $n$-qubit lattice Hamiltonian for time $t$, our algorithm has circuit depth $n^{o(1)}t^{1+o(1)}$. As a side application, our analysis gives a tensor network representation of lattice systems with bond dimension $2^{n^{o(1)}t^{1+o(1)}}$, using the counting argument of \cite{Jozsa06}. This significantly improves a recent construction of \cite[Lemma 17]{HP17} which uses only the first-order Trotter decomposition.

We work primarily with an idealized setting where quantum operations can be performed faithfully. However, in realistic experiments, quantum gates will be imperfectly implemented \cite{EZLBY18}. For such a case, Reference \cite{KM15} estimates the simulation accuracy as $\frac{\alpha}{r^{2k}}+\beta r$ in diamond-norm distance \cite{bk:wat,bk:wilde}, where $\alpha$ captures the algorithmic error of product formulas and $\beta$ captures gate errors. This leads to an optimal number of segments $r=\big(\alpha 2k/\beta\big)^{\frac{1}{2k+1}}$, which can be improved by our result. Specifically, the original analysis in \cite{BACS05} implies $\alpha_{\text{orig}}=O\big((nt)^{2k+1}\big)$. This has been improved by subsequent work \cite{CMNRS18,COS18}, although none of these improvements achieves linear scaling in $n$. In contrast, the analysis of this letter gives $\alpha_{\text{opt}}=O\big(nt^{2k+1}\big)$, improving the performance as a function of $n$ even in the presence of noise.

\begin{figure}[t]
	\centering
	\begin{subfigure}{.75\linewidth}
		\resizebox{.95\textwidth}{!}{
			\includegraphics{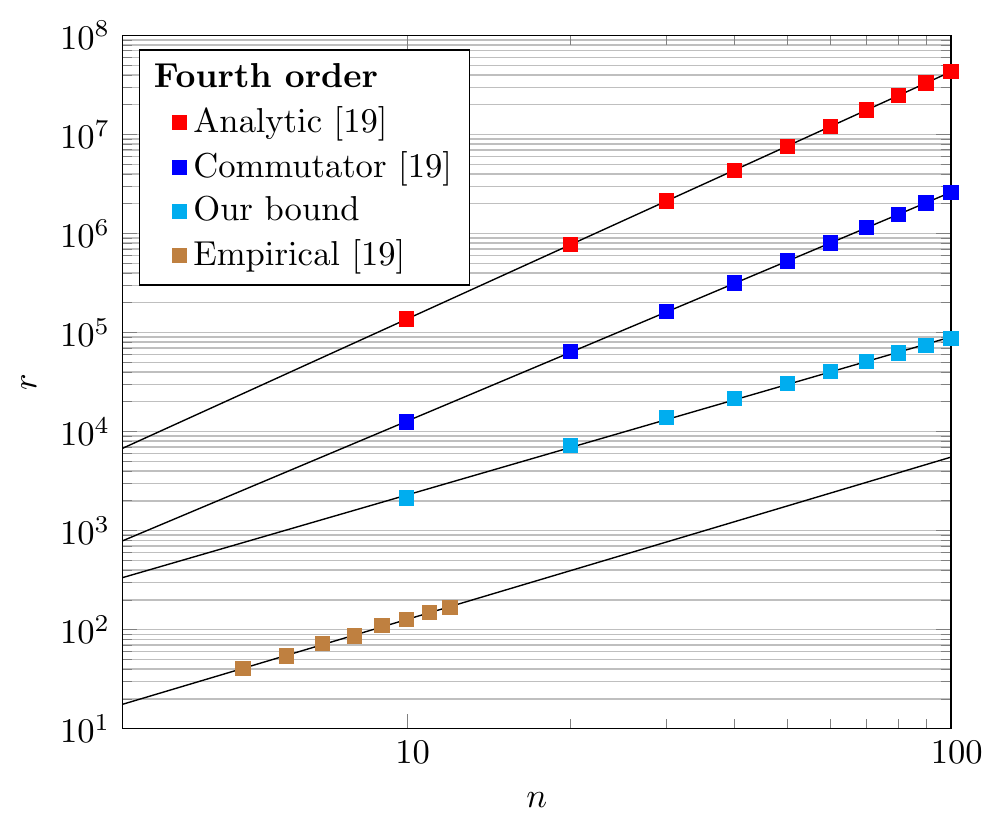}
		}
	\end{subfigure}%
	\caption{ Comparison of $r$ for different product-formula bounds for the Heisenberg model (see \cite[Section VI]{CSsupp19} for detailed parameters). Error bars are omitted as they are negligibly small on the plot. Straight lines show power-law fits to the data. \label{fig:pfbound}}
\end{figure}

Our main goal is to establish the gate complexity of $(nt)^{1+o(1)}$ for the product-formula algorithm. However, our analysis is not only nearly optimal in the asymptotic regime but also appears to be much tighter in practice. For concreteness, we numerically implement and optimize our fourth-order bound, and compare it with previous product-formula analysis, for simulation of a one-dimensional Heisenberg model with a random magnetic field with open boundary conditions \cite[Eq.(98)]{CSsupp19} (see \fig{pfbound}). We find that the scaling of our bound matches the empirical performance and the constant prefactor is off by only one order of magnitude, a significant improvement over previous rigorous bounds \cite{CMNRS18}. Further improvements of our bound are possible by optimizing its numerical implementation; we leave a detailed study for future work \cite{CSTWWZ19}.

\medskip
\noindent\textbf{Analysis of the first-order algorithm.}
The key technique behind our approach is an integral representation of the error $\scrS(t)-e^{-itH}$ that we develop based on Descombes and Thalhammer's \emph{local error analysis} of product formulas \cite{DT10}. In the local error representation, the integrand is expressed as a linear combination of commutators nested with unitary conjugations, where the numbers of summands and nesting layers are both independent of $n$ and $t$. We use this representation to get the correct asymptotic gate count as a function of $n$ and $t$. In contrast, the conventional approach uses the Baker-Campbell-Hausdorff formula or naive Taylor expansion, which requires the manipulation of infinite series and appears to be technically challenging to analyze \cite{HJ18} \cite[Section I]{CSsupp19}.

To illustrate the proof idea, we show how to obtain $\norm{\scrS_1(t)-e^{-itH}}=O(nt^2)$ for the first-order formula. We differentiate $\scrS_1(t)$ and obtain
\begin{equation}
\begin{aligned}
\scrS_1'(t)
&=-iH\scrS_1(t)+\big[e^{-itH_{\text{even}}},\ -iH_{\text{odd}}\big]e^{-itH_{\text{odd}}}.
\end{aligned}
\end{equation}
Using the variation-of-constants formula \cite{DT10} {{\cite[Theorem 4.9]{bk:knapp}}} with initial condition $\scrS_1(0)=I$, we find an integral representation of the product-formula error $\scrS_1(t)-e^{-itH}$ as
\begin{equation}
\int_{0}^{t}\mathrm{d}\tau_1\ e^{-i(t-\tau_1)H}\big[e^{-i\tau_1H_{\text{even}}},\ -iH_{\text{odd}}\big]e^{-i\tau_1H_{\text{odd}}}.
\end{equation}
We repeat this procedure to analyze the commutator $\big[e^{-i\tau_1H_{\text{even}}},\ -iH_{\text{odd}}\big]$, obtaining an upper bound on the spectral-norm error
\begin{equation}
\label{eq:local_error_pf1}
\begin{aligned}
\norm{\scrS_1(t)-e^{-itH}}
\leq\int_{0}^{t}\mathrm{d}\tau_1 \int_{0}^{\tau_1}\mathrm{d}\tau_2
\norm{\big[H_{\text{even}},\ H_{\text{odd}}\big]}.
\end{aligned}
\end{equation}

We expand $H_{\text{odd}}$ and $H_{\text{even}}$ according to their definitions. Fixing an arbitrary term $H_{2k-1,2k}$ in $H_{\text{odd}}$, the commutator $\big[H_{2l,2l+1},\ H_{2k-1,2k}\big]$ is non-zero only when $l\in\{k-1,k\}$. We thus find that
\begin{equation}
\label{eq:commutator_overlap}
\begin{aligned}
\big[H_{\text{even}},\ H_{\text{odd}}\big]
=\sum_{k=1}^{\frac{n}{2}}\big[H_{2k-2,2k-1}+H_{2k,2k+1},\ H_{2k-1,2k}\big].
\end{aligned}
\end{equation}
Using the triangle inequality, we have $\norm{\scrS_1(t)-e^{-itH}}=O(nt^2)$, which proves the claim \eq{main_result} for the first-order case.

\medskip
\noindent\textbf{Ordering robustness.}
Our above bound works when terms of the lattice Hamiltonian are ordered in the even-odd pattern. However, this choice is not necessary: the first-order algorithm has the same asymptotic error bound with respect to any ordering of the lattice terms.

Our analysis relies on an error bound for swapping lattice terms:
\begin{equation}
\label{eq:swap}
\begin{aligned}
\norm{\big[e^{-itH_{k,k+1}},e^{-itH_{l,l+1}}\big]}
\leq
2t^2
\end{aligned}
\end{equation}
if $|k-l|=1$ and $0$ otherwise. In words, we may swap two exponentials $e^{-itH_{k,k+1}}$ and $e^{-itH_{l,l+1}}$ without error unless their supports overlap, in which case the error is $O(t^2)$.

Let $H=\sum_{j=1}^{n-1}H_{j,j+1}$ be a lattice Hamiltonian. We now simulate it using the first-order formula, but allow terms to be ordered arbitrarily as $\prod_{j=1}^{n-1}e^{-itH_{\sigma(j),\sigma(j)+1}}$, where $\sigma\in S_{n-1}$ is a permutation on the $n-1$ elements $\{1,\ldots,n-1\}$. Then the spectral-norm error $\norm{\prod_{j=1}^{n-1}e^{-itH_{\sigma(j),\sigma(j)+1}}-e^{-itH}}$ is upper bounded by
\begin{equation}
\begin{aligned}
&\norm{\prod_{j=1}^{n-1}e^{-itH_{\sigma(j),\sigma(j)+1}}-e^{-itH_{\text{even}}}e^{-itH_{\text{odd}}}}\\
&+\norm{e^{-itH_{\text{even}}}e^{-itH_{\text{odd}}}-e^{-itH}}.
\end{aligned}
\end{equation}
The second term is upper bounded by $O(nt^2)$. For the first term, we transform $\prod_{j=1}^{n-1}e^{-itH_{\sigma(j),\sigma(j)+1}}$ into $\prod_{k=1}^{\frac{n}{2}-1}e^{-it H_{2k,2k+1}}\prod_{k=1}^{\frac{n}{2}}e^{-itH_{2k-1,2k}}$ by swapping neighboring exponentials. Every time two exponentials are swapped, we use \eq{swap} to bound the error. The total number of swaps of exponentials $e^{-itH_{k,k+1}}$ and $e^{-itH_{l,l+1}}$ with $|k-l|=1$ is at most $2n$, incurring error $4nt^2=O(nt^2)$.

We have therefore obtained the same asymptotic error bound for an arbitrary ordering of the Hamiltonian terms. We call this phenomenon \emph{ordering robustness}. Our analysis shows that the first-order algorithm is ordering-robust. Whether a similar property holds for a general higher-order formula remains an open question.

We also numerically compare the first-order algorithm with the even-odd ordering and the ordering of \cite{CMNRS18}. Although they have the same asymptotic error bound, in practice the even-odd ordering has smaller exponent and constant prefactor. Details can be found in \cite[Section VI]{CSsupp19}.

\medskip
\noindent\textbf{Analysis of higher-order algorithms.}
Analyzing higher-order product formulas is more challenging. To this end, we represent them in a \emph{canonical form}, which is easy to manipulate and encompasses well-known constructions such as the Lie-Trotter-Suzuki formulas $\scrS_1(t)$, $\scrS_{2k}(t)$ as special cases. We then use the variation-of-constants formula to write
\begin{equation}
\label{eq:local_R}
\begin{aligned}
\scrS_{2k}(t)-e^{-itH}=\int_{0}^{t} e^{-i(t-\tau)H}\mathscr{S}_{2k}(\tau)\mathscr{T}(\tau)\mathrm{d}\tau,
\end{aligned}
\end{equation}
where $\scrT(t)=\mathscr{S}_{2k}^\dagger(t)\big[\frac{\mathrm{d}}{\mathrm{d}t}\scrS_{2k}(t)-(-iH)\scrS_{2k}(t)\big]$. As a ($2k$)th-order formula, $\scrS_{2k}(t)$ satisfies an \emph{order condition} $\scrS_{2k}(t)=e^{-itH}+O(t^{2k+1})$, which further implies by Taylor's theorem
\begin{equation}
\label{eq:local_p_T}
\mathscr{T}(\tau)
=2k\int_{0}^{1}\mathrm{d}x\ (1-x)^{2k-1}\mathscr{T}^{(2k)}(x\tau)\frac{\tau^{2k}}{(2k)!}.
\end{equation}
Canonical product formulas and their order conditions are further discussed in \cite[Section II]{CSsupp19}.

A direct expansion of $\scrT(t)$ gives the correct $t$-dependence $O(t^{2k+1})$ of the product-formula error, but the scaling in $n$ is incorrect. Instead, we seek an alternative expression for the integrand that consists of a linear combination of commutators nested with unitary conjugations, where the number of summands and nested layers are both independent of $n$ and $t$. Such an expression is referred to as a \emph{local error representation} in \cite{DT10}. However, the result of \cite{DT10} depends on auxiliary functions whose recursive structure is hard to unravel. Instead, we develop a simpler representation of the local error structure \cite[Section III]{CSsupp19}.

In our representation, the operator $\mathscr{T}(\tau)$ can be written as a linear combination of operator-valued functions of the form $e^{i\tau X_1}\cdots e^{i\tau X_l}Ye^{-i\tau X_l}\cdots e^{-i\tau X_1}$, where operators $X_j, Y\in\{H_{\text{even}},H_{\text{odd}}\}$. As such, its higher-order derivatives consist of unitary conjugations and commutators. When a commutator is composed, we imitate \eq{commutator_overlap} to show that the support of the operator is expanded by at most a constant factor. When a unitary conjugation is composed, we decompose the unitary operators and cancel exponentials with non-overlapping supports. Throughout this procedure, we only introduce $O(n)$ error in the innermost layer, proving the claim in \eqref{eq:main_result} for the higher-order cases. This error analysis is discussed in more details in \cite[Section IV]{CSsupp19}.

\medskip
\noindent\textbf{Generalized lattice Hamiltonians.}
We have so far focused on time-independent one-dimensional systems with nearest-neighbor interactions and open boundary conditions. However, our analysis can be easily adapted to handle time-dependent Hamiltonians, periodic boundary conditions, constant-range interactions, and higher-dimensional systems, again with nearly optimal gate complexity.

When the Hamiltonian $H(t)$ is time-dependent, the problem of quantum simulation becomes more difficult \cite{WBHS10}. Then there no longer exists a closed-form solution to the Schr\"{o}dinger equation. Furthermore, some quantum simulation algorithms \cite{BC12,LC17} that behave well in the time-independent case fail to handle the time-dependent Hamiltonian simulation. Nevertheless, we show that product formulas can simulate time-dependent lattice Hamiltonians with nearly optimal gate complexity. We group the terms in the even-odd pattern
\begin{equation}
\begin{aligned}
H_{\text{odd}}(t)&=H_{1,2}(t)+H_{3,4}(t)+\cdots\\
H_{\text{even}}(t)&=H_{2,3}(t)+H_{4,5}(t)+\cdots
\end{aligned}
\end{equation}
and simulate using the time-dependent Lie-Trotter-Suzuki formulas $\scrS_{\mathcal{T},2k}(t)$ \cite{WBHS10}. We show that
\begin{equation}
\norm{\scrS_{\mathcal{T},2k}(t)-\exp_{\mathcal{T}}\bigg({-}i\int_{0}^{t}\mathrm{d}v\ H(v)\bigg)}=O\big(nt^{2k+1}\big)
\end{equation}
where $\exp_{\mathcal{T}}$ denotes the time-ordered matrix exponential. Similar to the time-independent case, we find that the total gate complexity is $O\big((nt)^{1+\frac{1}{2k}}\big)$. See \cite[Section V]{CSsupp19} for detailed discussions.

We also consider lattice Hamiltonians with periodic boundary conditions $H=\sum_{j=1}^{n-1}H_{j,j+1}+H_{1,n}$, where $H_{j,k}$ represents a local term that acts nontrivially only on qubits $j$ and $k$. To simulate such a system, we decompose $H$ as $H=H_{\text{bndry}}+H_{\text{even}}+H_{\text{odd}}$, where $H_{\text{bndry}}=H_{1,n}$. Correspondingly, we also use a canonical product formula with three exponentials per stage. With a similar analysis for the open boundary condition, we find that the product-formula error is $O(nt^{2k+1})$ as expected.

A generalization of this approach allows us to simulate a $D$-dimensional lattice Hamiltonian with nearly optimal gate complexity. We use a $2D$-coloring of the edges of the lattice to decompose the Hamiltonian into $2D$ terms, each of which is a sum of commuting terms. We also extend the definition of canonical product formulas to allow for $2D$ exponentials per stage. An analysis of the local error structure shows that this algorithm has gate complexity $O((L^D t)^{1+\frac{1}{2k}}/\epsilon^{\frac{1}{2k}})=O((nt)^{1+\frac{1}{2k}}/\epsilon^{\frac{1}{2k}})$, where $n$ is the total number of lattice sites and $L=n^{\frac{1}{D}}$ is the linear size of the lattice.

Finally, our algorithm can also simulate lattice Hamiltonians with constant-range interactions $H=\sum_{j=1}^{n-\ell+1}H_{j,\ldots,j+\ell-1}$. To achieve nearly-optimal gate complexity, we classify the Hamiltonian terms into the $\ell$ groups
\begin{equation}
\begin{aligned}
H_{[1]}=&\ H_{1,\ldots,\ell}+H_{\ell+1,\ldots,2\ell}+\cdots\\
H_{[2]}=&\ H_{2,\ldots,\ell+1}+H_{\ell+2,\ldots,2\ell+1}+\cdots\\
\vdots\ &\\
H_{[\ell]}=&\ H_{\ell,\ldots,2\ell-1}+H_{2\ell,\ldots,3\ell-1}+\cdots
\end{aligned}
\end{equation}
and use a product formula with $\ell$ elementary exponentials per stage.

\medskip
\noindent\textbf{Discussion.}
The product-formula algorithm is arguably the simplest approach to quantum simulation. We have showed that this approach can simulate lattice Hamiltonians with nearly optimal gate complexity. Our algorithm invokes product formulas by ordering terms in an even-odd pattern, which is conceptually easy to understand and straightforward to implement. Beyond the one-dimensional time-independent system with nearest-neighbor interactions and open boundary conditions, our analysis is also applicable to periodic boundary conditions, constant-range interactions, higher dimensions, and the time-dependent case, all with nearly optimal gate complexity. Our result also gives product-formula bounds that are much tighter in practice.

Recently, Haah, Hastings, Kothari and Low (HHKL) proposed another nearly optimal algorithm for lattice simulation \cite{HHKL18}. Instead of analyzing the product-formula approach, they develop a new approach motivated by the Lieb-Robinson bound \cite{Osborne06,Hastings10}, which quantifies how fast information can propagate in a system with local interactions. HHKL decomposes the entire evolution into blocks, where each block involves forward and backward evolution on a small region. Using product formulas within each block, their approach gives an ancilla-free algorithm for lattice simulation with asymptotic gate complexity $(nt)^{1+o(1)}$. However, this results in a much larger constant prefactor in practice than the pure product-formula algorithm analyzed here \cite[Section VI]{CSsupp19}.

The near optimality of HHKL depends essentially on the use of a Lieb-Robinson bound. As noted in \cite{HHKL18}, it may be difficult to apply this idea to Hamiltonians whose interactions are described by general graphs. Our approach directly exploits the commutation of lattice terms without the help of Lieb-Robinson bounds, which we expect could illuminate the simulation of other physical systems \cite{BWMMNC18,Tran18,ME11,PHOW19}.

Our local error analysis represents the product-formula error as an integral of a linear combination of commutators nested with unitary conjugations. Similar techniques have been used to establish the Lieb-Robinson bound and to study computational complexity aspects of many-body physics \cite{Osborne06,Hastings10,Tran18,AAVL11}. We leave it as an avenue for future work to explore whether these techniques could find more applications in the study of locality in quantum systems.

\section*{Acknowledgments}

\begin{acknowledgements}

Y.S. thanks Nathan Wiebe, Guang Hao Low, Minh Cong Tran, Stephen Jordan, Jeongwan Haah, Robin Kothari, Su-Kuan Chu, Rolando Somma, Xin Wang, Alexey Gorshkov, James R. Garrison, Anurag Anshu, Xiaodi Wu, Leonard Wossnig, Salini Karuvade, Xiao Yuan, Simon Benjamin, Shuchen Zhu, Scott Lawrence, and Zohreh Davoudi for helpful discussions. We thank anonymous referees for their comments. This work was supported in part by the Army Research Office (MURI award W911NF-16-1-0349), the Canadian Institute for Advanced Research, the National Science Foundation (grants 1526380 and 1813814), and the U.S.\ Department of Energy, Office of Science, Office of Advanced Scientific Computing Research, Quantum Algorithms Teams and Quantum Testbed Pathfinder programs.

\end{acknowledgements}

\bibliography{PFLatticeSim}

\pagebreak
\clearpage


\section*{Supplementary material}

# Nearly optimal lattice simulation by product formulas – Supplementary Material

Andrew M. Childs and Yuan Su

*Department of Computer Science, Institute for Advanced Computer Studies, and Joint Center for Quantum Information and Computer Science, University of Maryland*

(Dated: December 17, 2019)

## I. NONRIGOROUS ERROR ANALYSIS BY BAKER-CAMPBELL-HAUSDORFF FORMULA

In this section, we review an approach to product-formula error analysis based on the Baker-Campbell-Hausdorff (BCH) formula [1]. We explain why this argument is difficult to formalize and how local error analysis overcomes the difficulty.

Let $H = \sum_{j=1}^{n-1} H_{j,j+1}$ be an $n$-qubit lattice Hamiltonian, so the ideal evolution under $H$ for time $t$ is given by $e^{-itH}$. We group terms in an even-odd pattern as $H_{\mathrm{odd}} := H_{1,2} + H_{3,4} + \cdots = \sum_{k=1}^{\frac{n}{2}} H_{2k-1,2k}$, $H_{\mathrm{even}} := H_{2,3} + H_{4,5} + \cdots = \sum_{k=1}^{\frac{n}{2}-1} H_{2k,2k+1}$. For simplicity, we only analyze the first-order product formula $e^{-itH_{\mathrm{even}}} e^{-itH_{\mathrm{odd}}}$, which approximates the ideal evolution with error

$$e^{-itH_{\mathrm{even}}} e^{-itH_{\mathrm{odd}}} - e^{-itH}. \tag{1}$$

Jordan, Lee, and Preskill analyzed the scaling of this product-formula error as follows [1]. They first apply the BCH formula to the product formula and rewrite

$$e^{-itH_{\mathrm{even}}} e^{-itH_{\mathrm{odd}}} = e^{-itH - \frac{t^2}{2}[H_{\mathrm{even}}, H_{\mathrm{odd}}] + i\frac{t^3}{12}[H_{\mathrm{even}}, [H_{\mathrm{even}}, H_{\mathrm{odd}}]] - i\frac{t^3}{12}[H_{\mathrm{even}}, [H_{\mathrm{odd}}, H_{\mathrm{even}}]] + \cdots}. \tag{2}$$

Expanding the Taylor series and ignoring all higher-order terms, they obtain

$$e^{-itH_{\mathrm{even}}} e^{-itH_{\mathrm{odd}}} \approx e^{-itH} - \frac{t^2}{2}[H_{\mathrm{even}}, H_{\mathrm{odd}}]. \tag{3}$$

They thus estimate

$$\left\| e^{-itH_{\mathrm{even}}} e^{-itH_{\mathrm{odd}}} - e^{-itH} \right\| \approx O\big( \|[H_{\mathrm{even}}, H_{\mathrm{odd}}]\| t^2 \big) = O(nt^2), \tag{4}$$

which is the desired error scaling for the first-order product formula [2, Eq.(3)].

To formalize this argument, we must also consider higher-order terms. For a $p$th-order term in the Taylor series, we would instead estimate the spectral norm of a nested commutator

$$\left\| [H_{\mathrm{odd}}[\cdots, [H_{\mathrm{even}}, H_{\mathrm{odd}}]\cdots]] \right\| t^p. \tag{5}$$

By locality, this commutator scales like $O(nt^p)$ as long as the number of nesting layers is constant. However, when the number of layers is larger than $n$, the scaling becomes $O(n^p t^p)$. The $n$-dependence is now superlinear, which does not provide the desired error scaling [2, Eq.(3)]. See [3, Appendix B] for further discussion of this issue and drawbacks of this approach.

In comparison, local error analysis gives

$$
\begin{aligned}
&e^{-itH_{\mathrm{even}}} e^{-itH_{\mathrm{odd}}} - e^{-itH} \\
&= \int_0^t \mathrm{d}\tau_1 \int_0^{\tau_1} \mathrm{d}\tau_2 \ e^{-i(t-\tau_1)H} e^{-i\tau_1 H_{\mathrm{even}}} e^{i\tau_2 H_{\mathrm{even}}} [-iH_{\mathrm{even}}, \ -iH_{\mathrm{odd}}] e^{-i\tau_2 H_{\mathrm{even}}} e^{-i\tau_1 H_{\mathrm{odd}}}.
\end{aligned}
\tag{6}
$$

By the triangle inequality, we have

$$\left\| e^{-itH_{\mathrm{even}}} e^{-itH_{\mathrm{odd}}} - e^{-itH} \right\| \leq \int_0^t \mathrm{d}\tau_1 \int_0^{\tau_1} \mathrm{d}\tau_2 \left\| [H_{\mathrm{even}}, H_{\mathrm{odd}}] \right\| = O\big( \|[H_{\mathrm{even}}, H_{\mathrm{odd}}]\| t^2 \big) = O(nt^2). \tag{7}$$

Similar to the analysis based on the BCH formula, we are effectively bounding the lowest-order error, but the analysis is now done in a fully rigorous way. Generalizations give similar (though more complicated) error expressions for higher-order product-formulas, which we discuss in detail in Section II, Section III, and Section IV.

## II. CANONICAL PRODUCT FORMULAS AND ORDER CONDITIONS

In this section, we introduce notation and terminology that is useful for studying higher-order formulas. Similar to the first-order case, it is instructive to study a setting where the Hamiltonian is the sum of two Hermitian terms $H = A + B$. The evolution of $H$ for time $t$ is given by the unitary operator $\mathscr{E}(t) = e^{-itH}$, which may then be simulated using a specific product-formula algorithm, such as the Lie-Trotter formula or the Suzuki formulas. We will not analyze these formulas case-by-case. Instead, we consider *canonical product formulas*, a universal concept that includes well-known constructions.

**Definition 1** (Canonical product formula). *Let $H$ be a Hamiltonian consisting of two terms $H = A + B$, where $A$ and $B$ are Hermitian operators. We say that an operator-valued function $\mathscr{S}(t)$ is a* canonical product formula *for $H = A + B$ if it has the form*

$$\mathscr{S}(t) := \mathscr{S}_s(t) \cdots \mathscr{S}_2(t) \mathscr{S}_1(t) = \left(e^{-itb_s B} e^{-ita_s A}\right) \cdots \left(e^{-itb_2 B} e^{-ita_2 A}\right) \left(e^{-itb_1 B} e^{-ita_1 A}\right), \tag{8}$$

*where $a_1, \ldots, a_s$ and $b_1, \ldots, b_s$ are real coefficients. The parameter $s$ denotes the number of stages, and $\mathscr{S}_j(t) = e^{-itb_j B} e^{-ita_j A}$ is the $j$th-stage operator for $j = 1, \ldots, s$. We let $u$ be an upper bound on the coefficients, i.e.,*

$$\max\{|a_1|, \ldots, |a_s|, |b_1|, \ldots, |b_s|\} \le u. \tag{9}$$

*Finally, we say that the product formula $\mathscr{S}(t)$ has order $p$ for some integer $p \ge 1$ if*

$$\mathscr{S}(t) = \mathscr{E}(t) + O(t^{p+1}). \tag{10}$$

*We call $\mathscr{S}(t)$ an $(s, p, u)$-formula if we need an explicit description of the parameters.*

Although common constructions of product formulas involve stages where exponentials can be ordered both as $e^{-itb_j B} e^{-ita_j A}$ and as $e^{-ita_j A} e^{-itb_j B}$, we can achieve such orderings by padding with identity operators. In particular, we now show in detail how some well-known constructions of product formulas can be recast in the canonical form.

**Example 1** (First-order formula). *The first-order formula $e^{-itB} e^{-itA}$ may be represented as a 1-stage canonical formula by setting $b_1 = a_1 = 1$, whereas its reversed version $e^{-itA} e^{-itB}$ is a 2-stage canonical formula with the choice $b_2 = 0, a_2 = b_1 = 1, a_1 = 0$.*

**Example 2** (Second-order formula). *The second-order formula $e^{-i\frac{t}{2}A} e^{-itB} e^{-i\frac{t}{2}A}$ may be represented as a 2-stage canonical formula by setting $b_2 = 0, a_2 = \frac{1}{2}, b_1 = 1, a_1 = \frac{1}{2}$, whereas its reversed version $e^{-i\frac{t}{2}B} e^{-itA} e^{-i\frac{t}{2}B}$ is a 2-stage canonical formula with the choice $b_2 = \frac{1}{2}, a_2 = 1, b_1 = \frac{1}{2}, a_1 = 0$.*

**Example 3** ((2k)th-order formula). *The (2k)th-order Suzuki formula $\mathscr{S}_{2k}(t)$ defined in [2, Eq.(2)] is an $(s, p, u)$-formula, where $s \le 2 \cdot 5^{k-1}$, $p = 2k$, and $u = 1$.*

We now study the order conditions of a product formula. (Similar order conditions are sketched in [4], but we discuss them here for completeness.) Whenever possible, we follow the notation and terminology of [5]. We need the following lemma.

**Lemma 1.** *Let $F(t)$ be an operator-valued function that is infinitely differentiable. Let $p \ge 1$ be a nonnegative integer. The following two conditions are equivalent.*

1. *Asymptotic scaling: $F(t) = O(t^{p+1})$.*

2. *Derivative condition: $F(0) = F'(0) = \cdots = F^{(p)}(0) = 0$.*

*Proof.* Condition 2 implies 1 by Taylor's theorem. Assuming Condition 1 holds, we must have that

$$\|F(t)\| \le C_1 t^{p+1} \tag{11}$$

for some $C_1 \ge 0$ (and for $t$ sufficiently small). Let $0 \le j \le p$ be the first integer such that $F^{(j)}(0) \ne 0$. We use Taylor's theorem to find $C_2 \ge 0$ such that

$$\|F(t)\| \ge \left\|F^{(j)}(0)\right\| \frac{t^j}{j!} - C_2 t^{j+1}. \tag{12}$$

We combine the above inequalities and divide both sides by $t^j$. Taking the limit $t \to 0$ gives us a contradiction. $\square$

By definition, a product formula $\mathscr{S}(t)$ has order $p$ for some integer $p \geq 1$ if

$$\mathscr{S}(t) = \mathscr{E}(t) + O(t^{p+1}) \tag{13}$$

holds for any $H = A + B$. Invoking Lemma 1, we find an equivalent order condition

$$\mathscr{S}^{(j)}(0) = (-iH)^j \tag{14}$$

for $0 \leq j \leq p$.

As in the first-order case, we seek an integral representation of the product-formula error $\mathscr{S}(t) - \mathscr{E}(t)$. To this end, we differentiate $\mathscr{S}(t)$ and rewrite the derivative as $\frac{\mathrm{d}}{\mathrm{d}t}\mathscr{S}(t) = (-iH)\mathscr{S}(t) + \mathscr{R}(t)$, where

$$\mathscr{R}(t) := \frac{\mathrm{d}}{\mathrm{d}t}\mathscr{S}(t) - (-iH).\mathscr{S}(t). \tag{15}$$

Recall that $\mathscr{S}(t)$ is accurate up to order $p \geq 1$. Therefore, $\mathscr{S}(0) = I$ and we obtain the initial value problem $\frac{\mathrm{d}}{\mathrm{d}t}\mathscr{S}(t) = (-iH)\mathscr{S}(t) + \mathscr{R}(t)$, $\mathscr{S}(0) = I$. The solution of this problem is given by the variation-of-constants formula

$$\mathscr{S}(t) - \mathscr{E}(t) = \int_0^t e^{-i(t-\tau)H}\mathscr{R}(\tau)\mathrm{d}\tau. \tag{16}$$

We now determine an order condition for the operator $\mathscr{R}(t)$. Since $\mathscr{S}(t)$ is at least first-order accurate, we have $\mathscr{S}^{(1)}(0) = -iH$ and therefore $\mathscr{R}(0) = \mathscr{S}^{(1)}(0) - (-iH).\mathscr{S}(0) = 0$. By taking derivatives iteratively, one can show that

$$\mathscr{R}(0) = \mathscr{R}^{(1)}(0) = \cdots = \mathscr{R}^{(p-1)}(0) = 0. \tag{17}$$

Conversely, if higher-order derivatives of $\mathscr{R}$ satisfy the above condition, we must have

$$\mathscr{S}^{(j)}(0) = (-iH)\mathscr{S}^{(j-1)}(0) = \cdots = (-iH)^j\mathscr{S}(0) \tag{18}$$

for $1 \leq j \leq p$. Using the fact that $\mathscr{S}(0) = I$, we have $\mathscr{S}^{(j)}(0) = (-iH)^j$ for $0 \leq j \leq p$. Therefore, our new order condition (17) is equivalent to (14).

We proceed to rewrite the integrand $\mathscr{R}$ using the product formula $\mathscr{S}$ and another operator. Specifically, we let $\mathscr{T}(t)$ be the operator such that

$$\mathscr{R}(t) = \mathscr{S}(t).\mathscr{T}(t). \tag{19}$$

In quantum simulation, the product formula $\mathscr{S}(t)$ is unitary and therefore $\mathscr{T}(t) = \mathscr{S}(t)^\dagger\mathscr{R}(t)$. However, we will see that $\mathscr{T}(t)$ has significantly richer structure than it might seem. Analyzing the combinatorial structure of $\mathscr{T}(t)$ will be the central topic of the next section. For now, we shall focus on its order condition.

We claim that (17) is equivalent to the order condition

$$\mathscr{T}^{(j)}(0) = 0 \qquad \text{for all } 0 \leq j \leq p-1. \tag{20}$$

By the general Leibniz rule

$$\mathscr{R}^{(j)}(0) = (\mathscr{S}\mathscr{T})^{(j)}(0) = \sum_{l=0}^{j}\binom{j}{l}\mathscr{S}^{(j-l)}(0)\mathscr{T}^{(l)}(0), \tag{21}$$

so (20) implies (17). We prove the converse by induction. For $j = 0$, we have $\mathscr{R}(0) = 0$ and $\mathscr{S}(0) = I$. Therefore, $\mathscr{R}(0) = \mathscr{S}(0)\mathscr{T}(0)$ implies that $\mathscr{T}(0) = 0$. Assume that $\mathscr{T}^{(l)}(0) = 0$ has been proved for all $0 \leq l \leq j \leq p-2$. We apply the general Leibniz rule to compute the $(j+1)$th-order derivative of $\mathscr{R}$ and find

$$0 = \mathscr{R}^{(j+1)}(0) = \sum_{l=0}^{j+1}\binom{j+1}{l}\mathscr{S}^{(j+1-l)}(0)\mathscr{T}^{(l)}(0) = \mathscr{S}(0)\mathscr{T}^{(j+1)}(0) = \mathscr{T}^{(j+1)}(0). \tag{22}$$

Therefore $\mathscr{T}^{(l)}(0) = 0$ for all $0 \leq l \leq j+1$.

We now summarize all the product-formula order conditions determined above in the following theorem.

**Theorem 1** (Order conditions for canonical product formulas). *Let $H$ be a Hamiltonian consisting of two terms $H = A + B$, where $A$ and $B$ are Hermitian operators. Let $p \geq 1$ be an integer and let $\mathscr{S}(t)$ be a canonical product formula for $H = A + B$. The following four conditions are equivalent.*

1. *$\mathscr{S}(t) = e^{-itH} + O(t^{p+1})$.*

2. *$\mathscr{S}^{(j)}(0) = (-iH)^j$ for all $0 \leq j \leq p$.*

3. *There is some infinitely differentiable operator-valued function $\mathscr{R}(t)$ with $\mathscr{R}^{(j)}(0) = 0$ for all $0 \leq j \leq p - 1$, such that*

$$\mathscr{S}(t) = e^{-itH} + \int_0^t e^{-i(t-\tau)H} \mathscr{R}(\tau) \mathrm{d}\tau. \tag{23}$$

4. *There is some infinitely differentiable operator-valued function $\mathscr{T}(t)$ with $\mathscr{T}^{(j)}(0) = 0$ for all $0 \leq j \leq p - 1$, such that*

$$\mathscr{S}(t) = e^{-itH} + \int_0^t e^{-i(t-\tau)H} \mathscr{S}(\tau) \mathscr{T}(\tau) \mathrm{d}\tau. \tag{24}$$

*Furthermore, the operator-valued functions $\mathscr{R}(t) = \frac{\mathrm{d}}{\mathrm{d}t} \mathscr{S}(t) - (-iH)\mathscr{S}(t)$ and $\mathscr{T}(t) = \mathscr{S}(t)^\dagger \mathscr{R}(t)$ are uniquely determined.*

*Proof.* We have already proved $1 \Leftrightarrow 2$, $2 \Rightarrow 3$, and $3 \Rightarrow 4$, except for the differentiability of $\mathscr{R}$ and $\mathscr{T}$, which follows trivially from the definitions $\mathscr{R}(t) = \frac{\mathrm{d}}{\mathrm{d}t} \mathscr{S}(t) - (-iH)\mathscr{S}(t)$ and $\mathscr{T}(t) = \mathscr{S}(t)^\dagger \mathscr{R}(t)$.

Assume Condition 3 holds for some $\mathscr{R}(t)$. Differentiation gives

$$\mathscr{S}'(t) = (-iH)e^{-itH} + (-iH)e^{-itH} \int_0^t e^{i\tau H} \mathscr{R}(\tau) \mathrm{d}\tau + e^{-itH} e^{itH} \mathscr{R}(t) = (-iH)\mathscr{S}(t) + \mathscr{R}(t). \tag{25}$$

Therefore, $\mathscr{R}(t) = \frac{\mathrm{d}}{\mathrm{d}t} \mathscr{S}(t) - (-iH)\mathscr{S}(t)$ is uniquely determined, and $3 \Rightarrow 2$ follows from our previous analysis. In a similar way, we can show that $\mathscr{T}(t) = \mathscr{S}(t)^\dagger \big(\frac{\mathrm{d}}{\mathrm{d}t} \mathscr{S}(t) - (-iH)\mathscr{S}(t)\big)$ is uniquely determined and $4 \Rightarrow 3$ thus follows. □

## III. SIMPLIFIED LOCAL ERROR REPRESENTATION

In Section II, we found an integral representation for the product-formula error $\mathscr{S}(t) - \mathscr{E}(t) = \int_0^t e^{-i(t-\tau)H} \mathscr{R}(\tau) \mathrm{d}\tau$. A direct Taylor expansion of $\mathscr{R}(t)$ gives the correct scaling in $t$ but an incorrect dependence on $n$. To address this issue, we introduced an auxiliary operator $\mathscr{T}(t)$.

A direct Taylor expansion of $\mathscr{T}(t)$ based on its definition $\mathscr{T}(t) = \mathscr{S}(t)^\dagger \mathscr{R}(t)$ does not give the correct $n$-dependence either. Instead, we construct an alternative expression for the integrand that consists of a linear combination of commutators nested with unitary conjugations, where the number of summands and the number of nested layers are both independent of $n$ and $t$. Such an expression is referred to as a *local error representation* in [5]. To this end, we compute $\mathscr{R}(t) = \frac{\mathrm{d}}{\mathrm{d}t} \mathscr{S}(t) - (-iH)\mathscr{S}(t)$ explicitly. We then perform unitary conjugation to create $\mathscr{S}(t)$ on the left-hand side of $\mathscr{R}(t)$. Correspondingly, the right-hand side will contain the desired expression for $\mathscr{T}(t)$.

Let $H$ be a Hamiltonian consisting of two terms $H = A + B$, so that the ideal evolution is given by $\mathscr{E}(t) = e^{-it(A+B)}$. Consider simulating this Hamiltonian using an $s$-stage higher-order formula written in the canonical form $\mathscr{S}(t) = \mathscr{S}_s(t) \cdots \mathscr{S}_2(t) \mathscr{S}_1(t)$, where $\mathscr{S}_j(t) = e^{-itb_j B} e^{-ita_j A}$ is the $j$th-stage operator and $a_1, \ldots, a_s$ and $b_1, \ldots, b_s$ are real numbers. We adopt the convention $\prod_{l=1}^s \mathscr{S}_l(t) = \mathscr{S}_s(t) \mathscr{S}_{s-1}(t) \cdots \mathscr{S}_1(t)$ and let $b_0 = 0$.

We define $\mathscr{R}(t) := \frac{\mathrm{d}}{\mathrm{d}t}\mathscr{S}(t) - (-iH)\mathscr{S}(t)$. Our goal is to obtain a concrete expression for $\mathscr{T}(t)$ satisfying $\mathscr{R}(t) = \mathscr{S}(t)\mathscr{T}(t)$. We have

$$
\begin{aligned}
\mathscr{R}(t) =& \frac{\mathrm{d}}{\mathrm{d}t}\left[\prod_{j=1}^{s}\mathscr{S}_j(t)\right] - (-i)(A+B)\prod_{j=1}^{s}\mathscr{S}_j(t) \\
=& \sum_{j=1}^{s}\left(\prod_{l=j+1}^{s}\mathscr{S}_l(t)\right)\left(\mathscr{S}_j(t)(-ia_jA) + (-ib_jB)\mathscr{S}_j(t)\right)\left(\prod_{l=1}^{j-1}\mathscr{S}_l(t)\right) \\
& - \sum_{j=1}^{s}\left((-ia_jA) + (-ib_jB)\right)\prod_{l=1}^{s}\mathscr{S}_l(t),
\end{aligned}
\tag{26}
$$

where the second equality follows from the rule of differentiation and the fact that $\mathscr{S}(t)$ is at least first-order accurate, so $\sum_{j=1}^{s}a_j = \sum_{j=1}^{s}b_j = 1$. We re-express the differences of operators as commutators to get

$$
\mathscr{R}(t) = -i\sum_{j=1}^{s}\left[\prod_{l=j}^{s}\mathscr{S}_l(t),\ a_jA + b_{j-1}B\right]\prod_{l=1}^{j-1}\mathscr{S}_l(t).
\tag{27}
$$

Performing the commutation sequentially, we have

$$
\mathscr{R}(t) = -i\sum_{j=1}^{s}\sum_{k=j}^{s}\left(\prod_{l=k+1}^{s}\mathscr{S}_l(t)\left[\mathscr{S}_k(t),\ a_jA + b_{j-1}B\right]\prod_{l=j}^{k-1}\mathscr{S}_l(t)\right)\prod_{l=1}^{j-1}\mathscr{S}_l(t).
\tag{28}
$$

To proceed, we interchange the order of summation, giving

$$
\begin{aligned}
\mathscr{R}(t) =& -i\sum_{k=1}^{s}\sum_{j=1}^{k}\prod_{l=k+1}^{s}\mathscr{S}_l(t)\left[\mathscr{S}_k(t),\ a_jA + b_{j-1}B\right]\prod_{l=1}^{k-1}\mathscr{S}_l(t) \\
=& -i\sum_{k=1}^{s}\prod_{l=k+1}^{s}\mathscr{S}_l(t)\left[\mathscr{S}_k(t),\ c_kA + d_{k-1}B\right]\prod_{l=1}^{k-1}\mathscr{S}_l(t),
\end{aligned}
\tag{29}
$$

where we define

$$
c_k := \sum_{j=1}^{k}a_j, \qquad d_k := \sum_{j=1}^{k}b_j.
\tag{30}
$$

Finally, we perform unitary conjugation to create $\mathscr{S}(t)$ on the left-hand side of (29). Specifically, we have

$$
\begin{aligned}
& -i\sum_{k=1}^{s}\prod_{l=k}^{s}\mathscr{S}_l(t)\cdot\left(c_kA + d_{k-1}B\right)\cdot\prod_{l=1}^{k-1}\mathscr{S}_l(t) \\
=& -i\mathscr{S}(t)\sum_{k=1}^{s}\prod_{l=k-1}^{1}\mathscr{S}_l^{\dagger}(t)\cdot\left(c_kA + d_{k-1}B\right)\cdot\prod_{l=1}^{k-1}\mathscr{S}_l(t)
\end{aligned}
\tag{31}
$$

and

$$
\begin{aligned}
& -i\sum_{k=1}^{s}\prod_{l=k+1}^{s}\mathscr{S}_l(t)\cdot\left(c_kA + d_{k-1}B\right)\cdot\prod_{l=1}^{k}\mathscr{S}_l(t) \\
=& -i\mathscr{S}(t)\sum_{k=1}^{s}\prod_{l=k}^{1}\mathscr{S}_l^{\dagger}(t)\cdot\left(c_kA + d_{k-1}B\right)\cdot\prod_{l=1}^{k}\mathscr{S}_l(t).
\end{aligned}
\tag{32}
$$

We have now established the following theorem.

**Theorem 2** (Simplified local error representation). *Let $H = A + B$ be a Hamiltonian, so that the ideal evolution induced by $H$ is $\mathscr{E}(t) = e^{-it(A+B)}$. Let $\mathscr{S}(t)$ be an s-stage product formula written in the canonical form*

$$\mathscr{S}(t) = \mathscr{S}_s(t) \cdots \mathscr{S}_2(t) \mathscr{S}_1(t) = \left(e^{-itb_s B} e^{-ita_s A}\right) \cdots \left(e^{-itb_2 B} e^{-ita_2 A}\right) \left(e^{-itb_1 B} e^{-ita_1 A}\right), \tag{33}$$

*where $a_1, \ldots, a_s$ and $b_1, \ldots, b_s$ are real numbers, and $\mathscr{S}_j(t) = e^{-itb_j B} e^{-ita_j A}$ is the j-th stage operator. Then the product-formula error $\mathscr{S}(t) - \mathscr{E}(t)$ admits the integral representation*

$$\mathscr{S}(t) - \mathscr{E}(t) = \int_0^t \mathscr{E}(t - \tau) \mathscr{R}(\tau) \mathrm{d}\tau, \tag{34}$$

*where*

$$\mathscr{R}(\tau) = \mathscr{S}(\tau) \mathscr{T}(\tau) \tag{35}$$

*and*

$$\begin{aligned}
\mathscr{T}(\tau) = -i \sum_{k=1}^{s} \Bigg( &\prod_{l=k-1}^{1} \mathscr{S}_l^\dagger(\tau) \cdot \left(c_k A + d_{k-1} B\right) \cdot \prod_{l=1}^{k-1} \mathscr{S}_l(\tau) \\
&- \prod_{l=k}^{1} \mathscr{S}_l^\dagger(\tau) \cdot \left(c_k A + d_{k-1} B\right) \cdot \prod_{l=1}^{k} \mathscr{S}_l(\tau) \Bigg).
\end{aligned} \tag{36}$$

*Furthermore, if $\mathscr{S}(t)$ is a pth-order product formula, then*

$$\mathscr{T}(\tau) = \int_0^\tau \mathrm{d}v \; \frac{\mathscr{T}^{(p)}(v)}{(p-1)!} (\tau - v)^{p-1} = p \int_0^1 \mathrm{d}x \; (1-x)^{p-1} \mathscr{T}^{(p)}(x\tau) \frac{\tau^p}{p!}. \tag{37}$$

*Proof.* Equation (36) follows from the discussion above. The integral representation (37) follows from Theorem 1 and Taylor's theorem with integral remainder. □

The local error representation developed by Descombes and Thalhammer [5, Theorem 1] is proved through a similar calculation as in (29), except that they use two additional rules for manipulating matrix exponentials: one for creating exponentials [5, Eq. (2.9a)] and the other for pushing matrix exponentials [5, Eq. (2.9b)]. Unfortunately, they overlooked a time-dependent term in their calculation when establishing the second rule. Furthermore, Descombes and Thalhammer's analysis relies on auxiliary functions defined recursively in terms of integrals denoted $\mathscr{I}_1$ and $\mathscr{I}_2$, whose combinatorial structure is hard to unravel. In contrast, our local error representation follows from a unitary conjugation trick that significantly simplifies the calculations. Therefore, we use our Theorem 2 in subsequent analysis of the product-formula algorithm.

## IV.  ADJOINT MAPPINGS AND ANALYSIS OF THE $p$TH-ORDER ALGORITHM

In this section, we give a detailed analysis of the $p$th-order product-formula algorithm for lattice simulation. We introduce the notion of adjoint mappings in Section IV A and use it to obtain a bound on the product-formula error in Section IV B.

### A.  Adjoint mappings

For any invertible matrix $X$, we define $\mathrm{Ad}_X$ to be the conjugation transformation given by

$$\mathrm{Ad}_X(Y) = XYX^{-1} \tag{38}$$

for any operator $Y$. Also for an arbitrary operator $X$, we define $\mathrm{ad}_X$ to be the commutator transformation, i.e.

$$\mathrm{ad}_X(Y) = [X, Y] = XY - YX \tag{39}$$

for any operator $Y$. These definitions are motivated by the notion of adjoint representation in the study of Lie groups and Lie algebras [6].

In the following proposition, we state a differentiation rule for Ad and ad, which will be useful when we compute the Taylor expansion of a multivariate function.

**Proposition 1** (Differentiation rule)**.** *Let $X$ be an operator and let $Y(t)$ be an operator-valued function that is infinitely differentiable. Then*

$$\frac{\mathrm{d}}{\mathrm{d}t}\big[\mathrm{Ad}_{e^{tX}}(Y(t))\big] = \mathrm{Ad}_{e^{tX}}(\mathrm{ad}_X(Y(t))) + \mathrm{Ad}_{e^{tX}}(Y'(t)). \tag{40}$$

*Proof.* The proof is a straightforward calculation. $\qquad\square$

**Corollary 1** (Higher-order differentiation rule)**.** *Let $p$ be a positive integer, let $X$ be an operator, and let $Y(t)$ be an operator-valued function that is infinitely differentiable. Then*

$$\frac{\mathrm{d}^p}{\mathrm{d}t^p}\big[\mathrm{Ad}_{e^{tX}}(Y(t))\big] = \sum_{j=0}^{p}\binom{p}{j}\mathrm{Ad}_{e^{tX}}\mathrm{ad}_X^j\big(Y^{(p-j)}(t)\big). \tag{41}$$

*Proof.* The claimed rule follows by Proposition 1 and the proof of the general Leibniz rule. $\qquad\square$

In our analysis, a sequence of operators of the form

$$\begin{aligned}
&\mathrm{Ad}_{e^{\tau_1}X_{11}}\cdots\mathrm{Ad}_{e^{\tau_1}X_{1\gamma_1}}\mathrm{ad}_{Y_{11}}\cdots\mathrm{ad}_{Y_{1\delta_1}}\\
&\mathrm{Ad}_{e^{\tau_1}X_{21}}\cdots\mathrm{Ad}_{e^{\tau_1}X_{2\gamma_2}}\mathrm{ad}_{Y_{21}}\cdots\mathrm{ad}_{Y_{2\delta_2}}\\
&\cdots\\
&\mathrm{Ad}_{e^{\tau_m}X_{m1}}\cdots\mathrm{Ad}_{e^{\tau_m}X_{m\gamma_m}}\mathrm{ad}_{Y_{m1}}\cdots\mathrm{ad}_{Y_{m\delta_m}}(Z)
\end{aligned} \tag{42}$$

will be abbreviated as

$$\mathrm{Ad}_{\tau_1}^{\gamma_1}\mathrm{ad}^{\delta_1}\cdots\mathrm{Ad}_{\tau_m}^{\gamma_m}\mathrm{ad}^{\delta_m}(Z). \tag{43}$$

In other words, we omit the information about the operators and only keep track of the time variables $\tau_1,\ldots,\tau_m$. The advantage of this abbreviation is illustrated in the following proposition.

**Proposition 2** (Differentiation rule for abbreviated adjoint representation)**.** *The following differentiation rule for the abbreviated adjoint representation holds:*

$$\begin{aligned}
&\frac{\partial^{w_1+\cdots+w_m}}{\partial\tau_1^{w_1}\cdots\partial\tau_m^{w_m}}\mathrm{Ad}_{\tau_1}^{\gamma_1}\mathrm{ad}^{\delta_1}\cdots\mathrm{Ad}_{\tau_m}^{\gamma_m}\mathrm{ad}^{\delta_\ell}(Z)\\
&=\sum_{v_{11}+\cdots+v_{1\gamma_1}=w_1}\binom{w_1}{v_{11}\ \cdots\ v_{1\gamma_1}}\big(\mathrm{Ad}_{\tau_1}\mathrm{ad}^{v_{11}}\big)\cdots\big(\mathrm{Ad}_{\tau_1}\mathrm{ad}^{v_{1\gamma_1}}\big)\mathrm{ad}^{\delta_1}\\
&\cdots\\
&\qquad\sum_{v_{m1}+\cdots+v_{m\gamma_1}=w_m}\binom{w_m}{v_{m1}\ \cdots\ v_{m\gamma_m}}\big(\mathrm{Ad}_{\tau_m}\mathrm{ad}^{v_{m1}}\big)\cdots\big(\mathrm{Ad}_{\tau_m}\mathrm{ad}^{v_{m\gamma_m}}\big)\mathrm{ad}^{\delta_m}(Z).
\end{aligned} \tag{44}$$

*Proof.* To prove the stated rule, it suffices to separate the time variables and prove that

$$\frac{\partial^{w_1}}{\partial\tau_1^{w_1}}\mathrm{Ad}_{\tau_1}^{\gamma_1}(Z) = \sum_{v_{11}+\cdots+v_{1\gamma_1}=w_1}\binom{w_1}{v_{11}\ \cdots\ v_{1\gamma_1}}\big(\mathrm{Ad}_{\tau_1}\mathrm{ad}^{v_{11}}\big)\cdots\big(\mathrm{Ad}_{\tau_1}\mathrm{ad}^{v_{1\gamma_1}}\big)(Z). \tag{45}$$

This follows by Corollary 1 and the proof of the multi-factor Leibniz rule. $\qquad\square$

## B. Error analysis of the $p$th-order algorithm

Suppose that we want to simulate a Hamiltonian $H$ consisting of two terms $H = A + B$ for time $t$, so that the ideal evolution is given by $\mathscr{E}(t) = e^{-it(A+B)}$. As mentioned in Section II, a higher-order product formula may be represented in the canonical form

$$\mathscr{S}(t) = \mathscr{S}_s(t) \cdots \mathscr{S}_2(t) \mathscr{S}_1(t) = \left( e^{-itb_s B} e^{-ita_s A} \right) \cdots \left( e^{-itb_2 B} e^{-ita_2 A} \right) \left( e^{-itb_1 B} e^{-ita_1 A} \right), \tag{46}$$

where $s$ is the number of stages and $a_1, \ldots, a_s, b_1, \ldots, b_s \in \mathbb{R}$. By Theorem 2, we know that the product-formula error $\mathscr{S}(t) - \mathscr{E}(t)$ admits the integral representation

$$\mathscr{S}(t) - \mathscr{E}(t) = \int_0^t \mathscr{E}(t - \tau) \mathscr{R}(\tau) \mathrm{d}\tau, \tag{47}$$
$$\mathscr{R}(\tau) = \mathscr{S}(\tau) \mathscr{T}(\tau),$$

where

$$\mathscr{T}(\tau) = -i \sum_{k=1}^{s} \left\{ \prod_{l=k-1}^{1} \mathscr{S}_l^{\dagger}(\tau) \left( c_k A + d_{k-1} B \right) \prod_{l=1}^{k-1} \mathscr{S}_l(\tau) \right.$$
$$\left. - \prod_{l=k}^{1} \mathscr{S}_l^{\dagger}(\tau) \left( c_k A + d_{k-1} B \right) \prod_{l=1}^{k} \mathscr{S}_l(\tau) \right\}. \tag{48}$$

Fix $1 \leq k \leq s$. We observe that the first term in (48) has the abbreviated adjoint representation

$$\mathrm{Ad}_{\tau}^{2(k-1)} \left( c_k A + d_{k-1} B \right). \tag{49}$$

To establish the scaling $O(t^{p+1})$, it suffices to show that the $\tau$-dependence of $\mathscr{T}$ is $O(\tau^p)$. From Theorem 1, we know that terms of order $p - 1$ or less will vanish, so we only need to compute the integral remainder of the Taylor expansion of each $\mathrm{Ad}_{\tau}^{2(k-1)} \left( c_k A + d_{k-1} B \right)$ at order $p$. In light of the chain rule, we apply the multivariate Taylor theorem and obtain the remainder

$$p \int_0^1 \mathrm{d}s \ (1-s)^{p-1} \left[ \mathrm{Ad}_{\tau}^{2(k-1)} \left( c_k A + d_{k-1} B \right) \right]^{(p)} \frac{\tau^p}{p!}$$
$$= p \int_0^1 \mathrm{d}s \ (1-s)^{p-1}$$
$$\sum_{w_1 + \cdots + w_{2(k-1)} = p} \frac{\partial^p}{\partial \tau_1^{w_1} \cdots \partial \tau_{2(k-1)}^{w_{2(k-1)}}} \mathrm{Ad}_{\tau_1} \cdots \mathrm{Ad}_{\tau_{2(k-1)}} \left( c_k A + d_{k-1} B \right) \frac{\tau_1^{w_1} \cdots \tau_{2(k-1)}^{w_{2(k-1)}}}{w_1! \cdots w_{2(k-1)}!} \tag{50}$$
$$= p \int_0^1 \mathrm{d}s \ (1-s)^{p-1}$$
$$\sum_{w_1 + \cdots + w_{2(k-1)} = p} \mathrm{Ad}_{\tau_1} \mathrm{ad}^{w_1} \cdots \mathrm{Ad}_{\tau_{2(k-1)}} \mathrm{ad}^{w_{2(k-1)}} \left( c_k A + d_{k-1} B \right) \frac{\tau_1^{w_1} \cdots \tau_{2(k-1)}^{w_{2(k-1)}}}{w_1! \cdots w_{2(k-1)}!},$$

where $\tau_1 = \cdots = \tau_{2(k-1)} = \tau$.

We assume that $\mathscr{S}(t)$ is an $(s, p, u)$-formula. To simulate an $n$-qubit lattice Hamiltonian $H = \sum_{j=1}^{n-1} H_{j,j+1}$, we instantiate

$$A = H_{\mathrm{odd}} = H_{1,2} + H_{3,4} + \cdots = \sum_{k=1}^{\frac{n}{2}} H_{2k-1,2k}$$
$$B = H_{\mathrm{even}} = H_{2,3} + H_{4,5} + \cdots = \sum_{k=1}^{\frac{n}{2}-1} H_{2k,2k+1}. \tag{51}$$

We claim that

$$\left\| \mathrm{Ad}_{\tau_1} \mathrm{ad}^{w_1} \cdots \mathrm{Ad}_{\tau_{2(k-1)}} \mathrm{ad}^{w_{2(k-1)}} \left( c_k A + d_{k-1} B \right) \right\| \le n(2k-1)^p (ku)(2u)^p. \tag{52}$$

To see this, first note that we have operator $c_k A + d_{k-1} B$ in the inner-most layer, which contains at most $n$ terms, each of which has spectral norm at most $ku$. Now, we fix a particular term $c_k H_{2\eta-1,2\eta}$ and study the abbreviated adjoint representation in (52). The spectral norm will increase by a factor of $2u$ every time an ad is composed, and will remain the same if an Ad is composed. The total number of ad's is $p$, explaining the factor $(2u)^p$ in (52).

The justification of the factor $(2k-1)^p$ is more difficult. At the beginning, we have the operator $c_k H_{2\eta-1,2\eta}$. When the first ad is applied, we have

$$\left[ -ib_{k-1} H_{\mathrm{even}}, c_k H_{2\eta-1,2\eta} \right], \tag{53}$$

which only contains two nonzero commutators

$$\left[ -ib_{k-1} H_{2\eta-2,2\eta-1}, c_k H_{2\eta-1,2\eta} \right], \qquad \left[ -ib_{k-1} H_{2\eta,2\eta+1}, c_k H_{2\eta-1,2\eta} \right]. \tag{54}$$

We see that the support of the operator is enlarged from qubits $2\eta - 1, 2\eta$ to $2\eta - 2, 2\eta - 1, 2\eta, 2\eta + 1$. The next $w_{2(k-1)} - 1$ ad's all represent commutators with $-ib_{k-1} H_{\mathrm{even}}$. Therefore, the support of the operator will remain unchanged if more ad's are composed. When the next Ad is composed, we break the exponential of $H_{\mathrm{even}}$ into product of elementary exponentials of $H_{2\eta,2\eta+1}$, and cancel as many terms as possible in pairs. This does not enlarge the support either.

However, the next ad represents a commutator with $-ia_{k-1} H_{\mathrm{odd}}$. After cancellation, the support of the operator is enlarged to

$$2\eta - 3, 2\eta - 2, 2\eta - 1, 2\eta, 2\eta + 1, 2\eta + 2. \tag{55}$$

Following this argument, we find that the support of operators increases by two every time an Ad is composed. The total number of Ad's is $2(k-1)$, so the support of the last operator will be at most $4k - 2$. This upper bounds the number of nonzero nested commutators by $(2k-1)^p$.

The analysis is similar when the term in the inner-most layer of (52) is $d_{k-1} H_{2\eta,2\eta+1}$. Therefore, the remainder is upper bounded by

$$\begin{aligned} & p \int_0^1 \mathrm{d}s \ (1-s)^{p-1} \sum_{w_1 + \cdots + w_{2(k-1)} = p} n(2k-1)^p (ku)(2u)^p \frac{\tau^p}{w_1! \cdots w_{2(k-1)}!} \\ =& n(2k-1)^p (ku)(2u)^p (2k-2)^p \frac{\tau^p}{p!}. \end{aligned} \tag{56}$$

A summation over $1 \le k \le s$ and an integration $\int_0^t \mathrm{d}\tau$ give

$$\sum_{k=1}^s n(2k-1)^p (ku)(2u)^p (2k-2)^p \frac{t^{p+1}}{(p+1)!}. \tag{57}$$

Similarly, we find that the second term in (48) can be upper bounded by

$$\sum_{k=1}^s n(2k+1)^p (ku)(2u)^p (2k)^p \frac{t^{p+1}}{(p+1)!}. \tag{58}$$

Therefore, the product-formula error scales like $O(nt^{p+1})$ assuming that $s$, $p$, and $u$ are constant.

## V. TIME-DEPENDENT PRODUCT FORMULAS AND LOCAL ERROR ANALYSIS

In this appendix, we discuss time-dependent product formulas and their local error analysis in detail. In Section V A, we introduce canonical formulas for time-dependent Hamiltonian simulation and study their order conditions. We then analyze the time-dependent local error structure in Section V B.

## A. Time-dependent canonical formulas and order conditions

Let $H(t)$ be a Hamiltonian that depends on the time variable $t$. We can express the evolution under $H(t)$ for time $t$ as

$$\mathscr{E}_{\mathcal{T}}(t) := \exp_{\mathcal{T}} \left( -i \int_0^t \mathrm{d}v \ H(v) \right), \tag{59}$$

where $\mathscr{E}_{\mathcal{T}}(t)$ denotes the time-ordered exponential. The operator $\mathscr{E}_{\mathcal{T}}(t)$ is unitary and satisfies the differentiation rule

$$\frac{\mathrm{d}}{\mathrm{d}t} \mathscr{E}_{\mathcal{T}}(t) = -iH(t)\mathscr{E}_{\mathcal{T}}(t). \tag{60}$$

Throughout this section, we assume that the Hamiltonian $H(t)$ and its terms are infinitely differentiable with respect to $t$, which ensures that a product formula can approximate the ideal evolution to the stated order. The infinite differentiability of $H(t)$ may be relaxed [7], but we impose this assumption to make the presentation cleaner.

Assuming that $H(t)$ is infinitely differentiable, we have the following rule for computing higher-order derivatives of $\mathscr{E}_{\mathcal{T}}(t)$.

**Lemma 2** (Higher-order derivatives of $\mathscr{E}_{\mathcal{T}}(t)$ [7, Lemma 1]). *Let $H(t)$ be a time-dependent Hamiltonian that is infinitely differentiable. Then the evolution operator $\mathscr{E}_{\mathcal{T}}(t) = \exp_{\mathcal{T}} \left( -i \int_0^t \mathrm{d}v \ H(v) \right)$ is also infinitely differentiable and its derivatives are*

$$\mathscr{E}_{\mathcal{T}}^{(j)}(t) = T_j(t)\mathscr{E}_{\mathcal{T}}(t), \tag{61}$$

*where the $T_j(t)$ are specified by the recurrence*

$$\begin{aligned} T_0 &= I, \\ T_j(t) &= -iT_{j-1}(t)H(t) + \frac{\mathrm{d}}{\mathrm{d}t}T_{j-1}(t). \end{aligned} \tag{62}$$

We now show that $T_j(t)$ satisfies the following higher-order recursive formula.

**Lemma 3** (Recursive formula for $T_j(t)$). *For all $j \in \mathbb{N}$, the operator-valued function $T_j(t)$ defined in (62) satisfies*

$$T_{j+1}(t) = -i \sum_{k=0}^{j} \binom{j}{k} H^{(k)}(t) T_{j-k}(t). \tag{63}$$

*Proof.* We first prove by induction that

$$\mathscr{E}_{\mathcal{T}}^{(j+1)}(t) = -i \sum_{k=0}^{j} \binom{j}{k} H^{(k)}(t) \mathscr{E}_{\mathcal{T}}^{(j-k)}(t). \tag{64}$$

For the base case $j = 0$, the claimed equality reduces to

$$\mathscr{E}_{\mathcal{T}}^{(1)}(t) = -iH^{(0)}(t)\mathscr{E}_{\mathcal{T}}^{(0)}(t), \tag{65}$$

which follows trivially from (60).

Now suppose that

$$\mathscr{E}_{\mathcal{T}}^{(j)}(t) = -i \sum_{k=0}^{j-1} \binom{j-1}{k} H^{(k)}(t) \mathscr{E}_{\mathcal{T}}^{(j-1-k)}(t). \tag{66}$$

Differentiating both sides of the above equation, we have

$$
\begin{aligned}
\mathscr{E}_{\mathcal{T}}^{(j+1)}(t) =& \frac{\mathrm{d}}{\mathrm{d}t} \mathscr{E}_{\mathcal{T}}^{(j)}(t) \\
=& -i \frac{\mathrm{d}}{\mathrm{d}t} \sum_{k=0}^{j-1} \binom{j-1}{k} H^{(k)}(t) \mathscr{E}_{\mathcal{T}}^{(j-1-k)}(t) \\
=& -i \sum_{k=0}^{j-1} \binom{j-1}{k} H^{(k+1)}(t) \mathscr{E}_{\mathcal{T}}^{(j-1-k)}(t) - i \sum_{k=0}^{j-1} \binom{j-1}{k} H^{(k)}(t) \mathscr{E}_{\mathcal{T}}^{(j-k)}(t) \\
=& -i \sum_{k=1}^{j} \binom{j-1}{k-1} H^{(k)}(t) \mathscr{E}_{\mathcal{T}}^{(j-k)}(t) - i \sum_{k=0}^{j-1} \binom{j-1}{k} H^{(k)}(t) \mathscr{E}_{\mathcal{T}}^{(j-k)}(t) \\
=& -i \sum_{k=0}^{j} \binom{j}{k} H^{(k)}(t) T_{j-k}(t).
\end{aligned}
\tag{67}
$$

Thus (64) follows by induction.

We now invoke Lemma 2 to find

$$
T_{j+1}(t) \mathscr{E}_{\mathcal{T}}(t) = \sum_{k=0}^{j} \binom{j}{k} [-i H^{(k)}(t)] T_{j-k}(t) \mathscr{E}_{\mathcal{T}}(t).
\tag{68}
$$

Canceling the unitary operator $\mathscr{E}_{\mathcal{T}}(t)$ proves the claimed recursive formula for $T_j(t)$. □

Now let $H(t)$ be a time-dependent Hamiltonian consisting of two terms $H(t) = A(t) + B(t)$, where $A(t)$ and $B(t)$ are Hermitian operators that are infinitely differentiable with respect to $t$. We simulate the evolution under $H(t)$ using a product formula of the form

$$
\begin{aligned}
\mathscr{S}_{\mathcal{T}}(t) =& \mathscr{S}_{\mathcal{T},s}(t) \cdots \mathscr{S}_{\mathcal{T},2}(t) \mathscr{S}_{\mathcal{T},1}(t) \\
=& \left( e^{-i t b_s B(t \beta_s)} e^{-i t a_s A(t \alpha_s)} \right) \cdots \left( e^{-i t b_2 B(t \beta_2)} e^{-i t a_2 A(t \alpha_2)} \right) \left( e^{-i t b_1 B(t \beta_1)} e^{-i t a_1 A(t \alpha_1)} \right),
\end{aligned}
\tag{69}
$$

where $a_k, b_k, \alpha_k, \beta_k$ are real numbers. We call $\mathscr{S}_{\mathcal{T}}(t)$ a time-dependent canonical product formula with $s$ stages, where $\mathscr{S}_{\mathcal{T},k}(t) = e^{-i t b_k B(t \beta_k)} e^{-i t a_k A(t \alpha_k)}$ denotes the $k$-th stage operator for $k = 1, \ldots, s$. Intuitively, this formula samples the Hamiltonian at times $t\beta_k, t\alpha_k$ and applies a time-independent product formula to approximate the ideal evolution.

We say that $\mathscr{S}_{\mathcal{T}}(t)$ is a $p$th-order product formula if

$$
\mathscr{S}_{\mathcal{T}}(t) = \mathscr{E}_{\mathcal{T}}(t) + O(t^{p+1}).
\tag{70}
$$

Using Lemma 1, we find the order condition $\mathscr{S}_{\mathcal{T}}^{(j)}(0) = \mathscr{E}_{\mathcal{T}}^{(j)}(0)$ for $0 \le j \le p$, which is equivalent to

$$
\mathscr{S}_{\mathcal{T}}^{(j)}(0) = T_j(0)
\tag{71}
$$

by Lemma 2.

Let $\mathscr{R}_{\mathcal{T}}(t)$ be an operator-valued function defined as

$$
\mathscr{R}_{\mathcal{T}}(t) := \frac{\mathrm{d}}{\mathrm{d}t} \mathscr{S}_{\mathcal{T}}(t) + i H(t) \mathscr{S}_{\mathcal{T}}(t).
\tag{72}
$$

Using the variation-of-parameters formula [8, Theorem 4.9], $\mathscr{R}_{\mathcal{T}}(t)$ facilitates an integral representation of the product-formula error:

$$
\mathscr{S}_{\mathcal{T}}(t) - \mathscr{E}_{\mathcal{T}}(t) = \exp_{\mathcal{T}} \left( -i \int_0^t \mathrm{d}v \ H(v) \right) \int_0^t \mathrm{d}\tau_1 \ \exp_{\mathcal{T}}^{\dagger} \left( -i \int_0^{\tau_1} \mathrm{d}v \ H(v) \right) \mathscr{R}_{\mathcal{T}}(\tau_1).
\tag{73}
$$

We claim that the order condition

$$
\mathscr{S}_{\mathcal{T}}^{(j)}(0) = T_j(0), \quad 0 \le j \le p
\tag{74}
$$

is equivalent to

$$\mathscr{R}_{\mathcal{T}}^{(j)}(0) = 0, \quad 0 \leq j \leq p-1. \tag{75}$$

To see this, first suppose that (74) holds. For $0 \leq j \leq p-1$, we have

$$
\begin{aligned}
\mathscr{R}_{\mathcal{T}}^{(j)}(0) &= \mathscr{S}_{\mathcal{T}}^{(j+1)}(0) - \sum_{k=0}^{j} \binom{j}{k} [-iH^{(k)}(0)] \mathscr{S}_{\mathcal{T}}^{(j-k)}(0) \\
&= T_{j+1}(0) - \sum_{k=0}^{j} \binom{j}{k} [-iH^{(k)}(0)] T_{j-k}(0) = 0,
\end{aligned}
\tag{76}
$$

where the first equality follows from the definition of $\mathscr{R}_{\mathcal{T}}$ and the general Leibniz rule; the second equality follows from the order condition (74); and the last equality follows from Lemma 3. Conversely, suppose that (75) holds. Using the general Leibniz rule, we have

$$\mathscr{S}_{\mathcal{T}}^{(j+1)}(0) = \sum_{k=0}^{j} \binom{j}{k} [-iH^{(k)}(0)] \mathscr{S}_{\mathcal{T}}^{(j-k)}(0) \tag{77}$$

for $0 \leq j \leq p-1$. But we know from Lemma 3 that $T_{j+1}$ satisfies the same recursive formula, with the base case

$$\mathscr{S}_{\mathcal{T}}(0) = I = T_0. \tag{78}$$

An inductive argument gives (74).

To get the correct scaling of the product-formula error in both $n$ and $t$, we introduce another operator-valued function $\mathscr{T}_{\mathcal{T}}$ defined as

$$\mathscr{T}_{\mathcal{T}} := \mathscr{S}_{\mathcal{T}}^{\dagger} \mathscr{R}_{\mathcal{T}}. \tag{79}$$

The order condition for $\mathscr{T}_{\mathcal{T}}$ is

$$\mathscr{T}_{\mathcal{T}}^{(j)}(0) = 0, \quad 0 \leq j \leq p-1, \tag{80}$$

which follows in the same way as in Section II. We have now established several order conditions for time-dependent Hamiltonian simulation, which we summarize as follows.

**Theorem 3** (Order conditions for time-dependent canonical product formulas). *Let $H(t)$ be a time-dependent Hamiltonian consisting of two terms $H(t) = A(t) + B(t)$, such that both $A(t)$ and $B(t)$ are infinitely differentiable. Let $p \geq 1$ be an integer and let $\mathscr{S}_{\mathcal{T}}(t)$ be a time-dependent canonical product formula for $H(t)$. The following four conditions are equivalent.*

1. *$\mathscr{S}_{\mathcal{T}}(t) = \exp_{\mathcal{T}}\left(-i\int_0^t \mathrm{d}v\ H(v)\right) + O(t^{p+1})$.*

2. *$\mathscr{S}_{\mathcal{T}}^{(j)}(0) = T_j(0)$ for all $0 \leq j \leq p$. Here, $T_j(t)$ are defined recursively as*

$$
\begin{aligned}
T_0 &= I, \\
T_j(t) &= T_{j-1}(t)\left[-iH(t)\right] + \frac{\mathrm{d}}{\mathrm{d}t} T_{j-1}(t).
\end{aligned}
\tag{81}
$$

3. *There is some infinitely differentiable operator-valued function $\mathscr{R}_{\mathcal{T}}(t)$ with $\mathscr{R}_{\mathcal{T}}^{(j)}(0) = 0$ for all $0 \leq j \leq p-1$, such that*

$$
\begin{aligned}
\mathscr{S}_{\mathcal{T}}(t) &= \exp_{\mathcal{T}}\left(-i\int_0^t \mathrm{d}v\ H(v)\right) \\
&\quad + \exp_{\mathcal{T}}\left(-i\int_0^t \mathrm{d}v\ H(v)\right) \int_0^t \mathrm{d}\tau_1\ \exp_{\mathcal{T}}^{\dagger}\left(-i\int_0^{\tau_1} \mathrm{d}v\ H(v)\right) \mathscr{R}_{\mathcal{T}}(\tau_1).
\end{aligned}
\tag{82}
$$

4. *There is some infinitely differentiable operator-valued function $\mathscr{T}_{\mathcal{T}}(t)$ with $\mathscr{T}_{\mathcal{T}}^{(j)}(0) = 0$ for all $0 \leq j \leq p - 1$, such that*

$$
\begin{aligned}
\mathscr{S}_{\mathcal{T}}(t) = {}& \exp_{\mathcal{T}}\left(-i\int_0^t \mathrm{d}v\; H(v)\right) \\
& + \exp_{\mathcal{T}}\left(-i\int_0^t \mathrm{d}v\; H(v)\right) \int_0^t \mathrm{d}\tau_1\; \exp_{\mathcal{T}}^{\dagger}\left(-i\int_0^{\tau_1} \mathrm{d}v\; H(v)\right) \mathscr{S}_{\mathcal{T}}(\tau_1) \mathscr{T}_{\mathcal{T}}(\tau_1).
\end{aligned}
\tag{83}
$$

*Furthermore, $\mathscr{R}_{\mathcal{T}}(t) = \frac{\mathrm{d}}{\mathrm{d}t}\mathscr{S}_{\mathcal{T}}(t) - [-iH(t)]\mathscr{S}_{\mathcal{T}}(t)$ and $\mathscr{T}_{\mathcal{T}}(t) = \mathscr{S}_{\mathcal{T}}(t)^{\dagger}\mathscr{R}_{\mathcal{T}}(t)$ are uniquely determined.*

## B. Time-dependent local error representation

We now derive a local error representation for time-dependent Hamiltonian simulation. Let $H(t)$ be a time-dependent Hamiltonian consisting of two terms $H(t) = A(t) + B(t)$. We assume that both $A(t)$ and $B(t)$ are infinitely differentiable, although this assumption can be relaxed using techniques from [7]. The ideal evolution under $H(t)$ for time $t$ is given by the time-ordered exponential

$$
\mathscr{E}_{\mathcal{T}}(t) = \exp_{\mathcal{T}}\left(-i\int_0^t \mathrm{d}v\; H(v)\right),
\tag{84}
$$

which we simulate using a time-dependent canonical product formula

$$
\begin{aligned}
\mathscr{S}_{\mathcal{T}}(t) = {}& \mathscr{S}_{\mathcal{T},s}(t) \cdots \mathscr{S}_{\mathcal{T},2}(t)\mathscr{S}_{\mathcal{T},1}(t) \\
= {}& \left(e^{-itb_s B(t\beta_s)} e^{-ita_s A(t\alpha_s)}\right) \cdots \left(e^{-itb_2 B(t\beta_2)} e^{-ita_2 A(t\alpha_2)}\right)\left(e^{-itb_1 B(t\beta_1)} e^{-ita_1 A(t\alpha_1)}\right).
\end{aligned}
\tag{85}
$$

We know from **Theorem 3** that the product-formula error admits an integral representation

$$
\mathscr{S}_{\mathcal{T}}(t) = \mathscr{E}_{\mathcal{T}}(t) + \mathscr{E}_{\mathcal{T}}(t) \int_0^t \mathrm{d}\tau_1\; \mathscr{E}_{\mathcal{T}}^{\dagger}(\tau_1)\mathscr{R}_{\mathcal{T}}(\tau_1),
\tag{86}
$$

where $\mathscr{R}_{\mathcal{T}}(t) = \frac{\mathrm{d}}{\mathrm{d}t}\mathscr{S}_{\mathcal{T}}(t) - [-iH(t)]\mathscr{S}_{\mathcal{T}}(t)$. A direct Taylor expansion of $\mathscr{R}_{\mathcal{T}}(t)$ will give the correct error scaling of $t$, but cannot easily be used to show the correct $n$-dependence. Instead, we consider an expansion of the operator $\mathscr{T}_{\mathcal{T}}(t) = \mathscr{S}_{\mathcal{T}}(t)^{\dagger}\mathscr{R}_{\mathcal{T}}(t)$. To this end, we compute $\mathscr{R}_{\mathcal{T}}(t) = \frac{\mathrm{d}}{\mathrm{d}t}\mathscr{S}_{\mathcal{T}}(t) - (-iH)\mathscr{S}_{\mathcal{T}}(t)$ explicitly. We then perform unitary conjugation to create $\mathscr{S}_{\mathcal{T}}(t)$ on the left-hand side of $\mathscr{R}_{\mathcal{T}}(t)$. Correspondingly, the right-hand side will contain the desired expression for $\mathscr{T}_{\mathcal{T}}(t)$.

The following lemma is useful in our analysis.

**Lemma 4** (Chain rule for matrix exponentiation [9, Eq.(29)], [10, Page 181])**.** *Let $G(x)$ be an operator-valued function of $x \in \mathbb{R}$ that is infinitely differentiable. Then the derivative $\frac{\mathrm{d}e^{G(x)}}{\mathrm{d}x}$ can be expressed as*

$$
\frac{\mathrm{d}e^{G(x)}}{\mathrm{d}x} = \int_0^1 e^{yG(x)}\frac{\mathrm{d}G(x)}{\mathrm{d}x}e^{(1-y)G(x)}\mathrm{d}y = \int_0^1 e^{(1-z)G(x)}\frac{\mathrm{d}G(x)}{\mathrm{d}x}e^{zG(x)}\mathrm{d}z.
\tag{87}
$$

*If we further define operator-valued functions*

$$
\begin{aligned}
\mathscr{I}_{\mathcal{T},\mathrm{L}}\big(G(x),x\big) &= \int_0^1 \mathrm{d}y\; e^{yG(x)}G'(x)e^{-yG(x)} \\
\mathscr{I}_{\mathcal{T},\mathrm{R}}\big(G(x),x\big) &= \int_0^1 \mathrm{d}y\; e^{-yG(x)}G'(x)e^{yG(x)},
\end{aligned}
\tag{88}
$$

*then the chain rule can be succinctly expressed as*

$$
\frac{\mathrm{d}e^{G(x)}}{\mathrm{d}x} = \mathscr{I}_{\mathcal{T},\mathrm{L}}\big(G(x),x\big)e^{G(x)} = e^{G(x)}\mathscr{I}_{\mathcal{T},\mathrm{R}}\big(G(x),x\big).
\tag{89}
$$

We now compute

$$
\begin{aligned}
\mathscr{R}_{\mathcal{T}}(t) =& \frac{\mathrm{d}}{\mathrm{d}t}\mathscr{S}_{\mathcal{T}}(t) - [-iH(t)]\mathscr{S}_{\mathcal{T}}(t) \\
=& \frac{\mathrm{d}}{\mathrm{d}t}\bigg[\prod_{j=1}^{s}\mathscr{S}_{\mathcal{T},j}(t)\bigg] + iH(t)\prod_{j=1}^{s}\mathscr{S}_{\mathcal{T},j}(t) \\
=& \sum_{j=1}^{s}\bigg(\prod_{l=j+1}^{s}\mathscr{S}_{\mathcal{T},l}(t)\bigg)\bigg(\mathscr{S}_{\mathcal{T},j}(t)\mathscr{I}_{\mathcal{T},\mathrm{R}}\big(-ita_j A(t\alpha_j), t\big) \\
& \hspace{4cm} + \mathscr{I}_{\mathcal{T},\mathrm{L}}\big(-itb_j B(t\beta_j), t\big)\mathscr{S}_{\mathcal{T},j}(t)\bigg)\bigg(\prod_{l=1}^{j-1}\mathscr{S}_{\mathcal{T},l}(t)\bigg) \\
& + iH(t)\prod_{l=1}^{s}\mathscr{S}_{\mathcal{T},l}(t),
\end{aligned}
\tag{90}
$$

where we have used the chain rule in the last equality. To proceed, we perform unitary conjugation to create the time-dependent product formula on the left-hand side as

$$
\begin{aligned}
& \sum_{j=1}^{s}\bigg(\prod_{l=j+1}^{s}\mathscr{S}_{\mathcal{T},l}(t)\bigg)\bigg(\mathscr{S}_{\mathcal{T},j}(t)\mathscr{I}_{\mathcal{T},\mathrm{R}}\big(-ita_j A(t\alpha_j), t\big) \\
& \hspace{4cm} + \mathscr{I}_{\mathcal{T},\mathrm{L}}\big(-itb_j B(t\beta_j), t\big)\mathscr{S}_{\mathcal{T},j}(t)\bigg)\bigg(\prod_{l=1}^{j-1}\mathscr{S}_{\mathcal{T},l}(t)\bigg) \\
=& \mathscr{S}_{\mathcal{T}}(t)\sum_{j=1}^{s}\bigg(\prod_{l=j-1}^{1}\mathscr{S}_{\mathcal{T},l}^{\dagger}(t)\bigg)\mathscr{I}_{\mathcal{T},\mathrm{R}}\big(-ita_j A(t\alpha_j), t\big)\bigg(\prod_{l=1}^{j-1}\mathscr{S}_{\mathcal{T},l}(t)\bigg) \\
& + \mathscr{S}_{\mathcal{T}}(t)\sum_{j=1}^{s}\bigg(\prod_{l=j}^{1}\mathscr{S}_{\mathcal{T},l}^{\dagger}(t)\bigg)\mathscr{I}_{\mathcal{T},\mathrm{L}}\big(-itb_j B(t\beta_j), t\big)\bigg(\prod_{l=1}^{j}\mathscr{S}_{\mathcal{T},l}(t)\bigg)
\end{aligned}
\tag{91}
$$

and

$$
\begin{aligned}
& iH(t)\prod_{l=1}^{s}\mathscr{S}_{\mathcal{T},l}(t) \\
=& \mathscr{S}_{\mathcal{T}}(t)\prod_{l=s}^{1}\mathscr{S}_{\mathcal{T},l}(t)\big[iH(t)\big]\prod_{l=1}^{s}\mathscr{S}_{\mathcal{T},l}(t).
\end{aligned}
\tag{92}
$$

We have therefore established the following.

**Theorem 4** (Time-dependent local error representation). *theorem Let* $H(t) = A(t) + B(t)$ *be a time-dependent Hamiltonian with* $A(t)$ *and* $B(t)$ *infinitely differentiable, so that the ideal evolution under* $H(t)$ *for time* $t$ *is given by* $\mathscr{E}_{\mathcal{T}}(t) = \exp_{\mathcal{T}}\big(-i\int_0^t \mathrm{d}v\ H(v)\big)$. *Let* $\mathscr{S}_{\mathcal{T}}(t)$ *be a time-dependent* $s$-*stage formula written in the canonical form*

$$
\begin{aligned}
\mathscr{S}_{\mathcal{T}}(t) =& \mathscr{S}_{\mathcal{T},s}(t)\cdots\mathscr{S}_{\mathcal{T},2}(t)\mathscr{S}_{\mathcal{T},1}(t) \\
=& \big(e^{-itb_s B(t\beta_s)}e^{-ita_s A(t\alpha_s)}\big)\cdots\big(e^{-itb_2 B(t\beta_2)}e^{-ita_2 A(t\alpha_2)}\big)\big(e^{-itb_1 B(t\beta_1)}e^{-ita_1 A(t\alpha_1)}\big),
\end{aligned}
\tag{93}
$$

*where* $a_k, b_k, \alpha_k, \beta_k$ *are real numbers and* $\mathscr{S}_{\mathcal{T},k}(t) = e^{-itb_k B(t\beta_k)}e^{-ita_k A(t\alpha_k)}$ *is the* $k$-*th stage operator for* $k \in \{1, \ldots, s\}$. *Then the product-formula error* $\mathscr{S}_{\mathcal{T}}(t) - \mathscr{E}_{\mathcal{T}}(t)$ *admits the integral representation*

$$
\begin{aligned}
\mathscr{S}_{\mathcal{T}}(t) - \mathscr{E}_{\mathcal{T}}(t) =& \int_0^t \mathscr{E}_{\mathcal{T}}(t-\tau)\mathscr{R}_{\mathcal{T}}(\tau)\mathrm{d}\tau, \\
\mathscr{R}_{\mathcal{T}}(\tau) =& \mathscr{S}_{\mathcal{T}}(\tau)\mathscr{T}_{\mathcal{T}}(\tau),
\end{aligned}
\tag{94}
$$

where

$$\mathcal{T}_{\mathcal{T}}(\tau) = \sum_{j=1}^{s} \left\{ \left( \prod_{l=j-1}^{1} \mathscr{S}_{\mathcal{T},l}^{\dagger}(t) \right) \mathscr{I}_{\mathcal{T},\mathrm{R}} \big( -ita_j A(t\alpha_j), t \big) \left( \prod_{l=1}^{j-1} \mathscr{S}_{\mathcal{T},l}(t) \right) \right.$$
$$\left. - \left( \prod_{l=j}^{1} \mathscr{S}_{\mathcal{T},l}^{\dagger}(t) \right) \mathscr{I}_{\mathcal{T},\mathrm{L}} \big( -itb_j B(t\beta_j), t \big) \left( \prod_{l=1}^{j} \mathscr{S}_{\mathcal{T},l}(t) \right) \right\}$$
$$+ \prod_{l=s}^{1} \mathscr{S}_{\mathcal{T},l}^{\dagger}(t) \big[ iH(t) \big] \prod_{l=1}^{s} \mathscr{S}_{\mathcal{T},l}(t) \tag{95}$$

and

$$\mathscr{I}_{\mathcal{T},\mathrm{L}} \big( G(x), x \big) = \int_0^1 \mathrm{d}y \ e^{yG(x)} G'(x) e^{-yG(x)}$$
$$\mathscr{I}_{\mathcal{T},\mathrm{R}} \big( G(x), x \big) = \int_0^1 \mathrm{d}y \ e^{-yG(x)} G'(x) e^{yG(x)}. \tag{96}$$

Furthermore, if $\mathscr{S}_{\mathcal{T}}(t)$ is a time-dependent pth-order formula, then

$$\mathcal{T}_{\mathcal{T}}(\tau) = p \int_0^1 \mathrm{d}x \ (1-x)^{p-1} \mathcal{T}_{\mathcal{T}}^{(p)}(x\tau) \frac{\tau^p}{p!}. \tag{97}$$

Here (97) follows from the order conditions and Taylor's theorem with integral remainder as in Theorem 2.

## VI. EMPIRICAL PERFORMANCE

The product-formula algorithm is the simplest approach to digital quantum simulation and its implementation does not require any ancilla qubits. We have shown that this algorithm can simulate a lattice Hamiltonian with nearly optimal gate complexity, and we established the ordering robustness property for the first-order algorithm. Recently, Haah, Hastings, Kothari, and Low (HHKL) proposed a new algorithm motivated by the Lieb-Robinson bounds, which also has nearly optimal complexity for lattice simulation and is ancilla-free if its each block is simulated by product formulas [11]. In this section, we numerically compare the empirical gate complexity of the product-formula algorithm and HHKL. We also consider the empirical performance of product formulas with respect to different orderings of lattice terms.

For concreteness, we consider a one-dimensional nearest-neighbor Heisenberg model with a random magnetic field. Its Hamiltonian has the form

$$H = \sum_{j=1}^{n-1} (\vec{\sigma}_j \cdot \vec{\sigma}_{j+1} + h_j \sigma_j^z) \tag{98}$$

with coefficients $h_j \in [-1, 1]$ chosen uniformly at random, where $\vec{\sigma}_j = (\sigma_j^x, \sigma_j^y, \sigma_j^z)$ denotes a vector of Pauli $x$, $y$, and $z$ matrices on qubit $j$. This is a widely studied model in condensed matter physics, whose simulation is beyond the reach of current classical computers for all but the smallest systems. Following [12], we set accuracy $\epsilon = 10^{-3}$ and choose the simulation time to be the same as the system size (i.e., $t = n$). To simplify the numerical implementation, we consider open boundary conditions, although our analysis can also be generalized to handle periodic conditions as described in [2]. This Hamiltonian has the form $H = 4 \sum_{j=1}^{n-1} H_{j,j+1}$ with $H_{j,j+1} = (\vec{\sigma}_j \cdot \vec{\sigma}_{j+1} + h_j \sigma_j^z)/4$, so that $\max_j \|H_{j,j+1}\| \leq 1$. Thus we normalize our Hamiltonian by a factor of 4 and simulate for time $4t$. We estimate the empirical gate complexity of product formulas as in [12].

In HHKL, the entire evolution is decomposed into $m/2$ blocks of evolutions on $\ell$ qubits and $m/2$ blocks of evolutions on $2\ell$ qubits, each for time $t_\square$. We choose $\ell = 7$ to be constant and use the fitted data of [11] to obtain an error contribution of

$$0.175 \left( \frac{7.9t_\square}{\ell + 0.95} \right)^{\ell + 0.95} = \frac{\epsilon}{3m}, \tag{99}$$

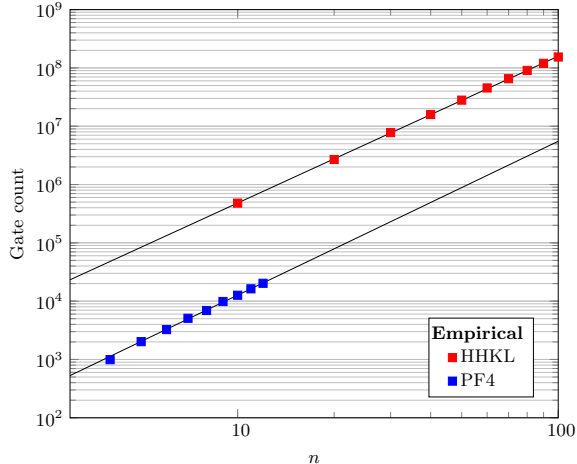

FIG. 1: Comparison of the empirical gate count between the HHKL algorithm with each block simulated by the fourth-order product formula and the (pure) fourth-order product-formula algorithm. Error bars are omitted as they are negligibly small on the plot. Straight lines show power-law fits to the data.

where the number of blocks is

$$m = \frac{8tn}{t_\square \ell}. \tag{100}$$

We can therefore simultaneously solve for the number of blocks $m$ and the evolution time per block $t_\square$. We then use product formulas to simulate each block of size $\ell = 7$ or $2\ell = 14$ for time $t_\square$ with accuracy $\epsilon_\square = \epsilon/3m$. We choose the fourth-order product formula since it has the minimum gate count in practice for simulating the Heisenberg model of size up to 30 [12, Figure 3]. For a fair comparison, we also compute the empirical gate count of the pure fourth-order product-formula algorithm, whose performance is not too much worse than the best product formula for $n$ up to 300 [12, Figure 3].

Figure 1 shows the resulting gate complexity for HHKL and the pure product-formula algorithm. Fitting the data, we obtain

$$g_{\text{HHKL}} = 1461.453n^{2.518}, \qquad g_{\text{PF}} = 29.093n^{2.639}. \tag{101}$$

We find that, while the asymptotic scaling of HHKL is better, the product-formula approach has a significantly better constant prefactor. Indeed, the HHKL algorithm introduces extra negative terms in the Hamiltonian to compensate for the error of the Lieb-Robinson decomposition and then simulates each block using product formulas [11], whereas the pure product-formula algorithm simulates the original lattice Hamiltonian with no overhead. Therefore, even though both algorithms are ancilla-free, the pure product-formula approach seems more desirable for near-term simulation.

To better understand the ordering robustness property, we also compare the empirical values of $r$ for the first-order product-formula algorithm by ordering terms in the even-odd pattern and the X-Y-Z pattern of [12]. Figure 2 shows the resulting data, from which we estimate

$$r_{\text{even-odd}} = 586.816n^{1.942}, \qquad r_{\text{X-Y-Z}} = 668.139n^{2.507}. \tag{102}$$

Both are consistent with the claimed upper bound $r = O(nt^2) = O(n^3)$ for lattice simulation, but the even-odd ordering of terms gives better performance in practice.

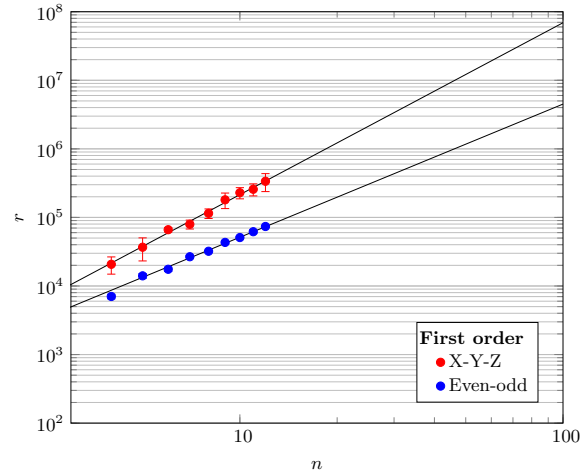

FIG. 2: Comparison of the empirical values of $r$ for the first-order product-formula algorithm with the even-odd ordering and the X-Y-Z ordering of [12]. Error bars are omitted when they are negligibly small on the plot. Straight lines show power-law fits to the data.



\end{document}